\documentclass[%
 reprint,
 amsmath,amssymb,
 aps,
 pre,
 floatfix,
]{revtex4-2}

\usepackage{graphicx}% Include figure files
\usepackage{dcolumn}% Align table columns on decimal point
\usepackage{bm}% bold math
\usepackage[colorlinks,linkcolor=blue,anchorcolor=blue,citecolor=blue,urlcolor=blue]{hyperref}

\usepackage{amsbsy} % math symbol bold
\usepackage{booktabs} % professional table lines

\begin{document}

\preprint{APS/123-QED}

\title{Frequency Locking to Environmental Forcing Suppresses Oscillatory Extinction in Phage-Bacteria Interactions}

\author{Hao-Neng Luo}
\affiliation{Lanzhou Center for Theoretical Physics, Key Laboratory of Theoretical Physics of Gansu Province, Key Laboratory of Quantum Theory and Applications of MoE, Gansu Provincial Research Center for Basic Disciplines of Quantum Physics, and Institute of Computational Physics and Complex Systems, Lanzhou University, Lanzhou 730000, China}

\author{Zhi-Xi Wu}\email{Contact author: wuzhx@lzu.edu.cn}
\affiliation{Lanzhou Center for Theoretical Physics, Key Laboratory of Theoretical Physics of Gansu Province, Key Laboratory of Quantum Theory and Applications of MoE, Gansu Provincial Research Center for Basic Disciplines of Quantum Physics, and Institute of Computational Physics and Complex Systems, Lanzhou University, Lanzhou 730000, China}

\author{Jian-Yue Guan}
\affiliation{Lanzhou Center for Theoretical Physics, Key Laboratory of Theoretical Physics of Gansu Province, Key Laboratory of Quantum Theory and Applications of MoE, Gansu Provincial Research Center for Basic Disciplines of Quantum Physics, and Institute of Computational Physics and Complex Systems, Lanzhou University, Lanzhou 730000, China}

\date{\today}

\begin{abstract}
Bacteriophage-bacteria interactions are central to microbial ecology, influencing evolution, biogeochemical cycles, and pathogen behavior. Most theoretical models assume static environments and passive bacterial hosts, neglecting the joint effects of bacterial traits and environmental fluctuations on coexistence dynamics. This limitation hinders the prediction of microbial persistence in dynamic ecosystems such as soils and oceans. Using a minimal ordinary differential equation framework, 
we demonstrate that environmental fluctuations can suppress destructive oscillations through resonance, promoting coexistence where static models otherwise predict collapse. Counterintuitively, we find that lower bacterial growth rates are helpful in enhancing survival under high infection pressure, elucidating the observed post-infection growth reduction. Our studies highlight bacterial hosts as active builders of ecological dynamics and environmental variation as a potential stabilizing force. Our findings thus bridge a theory-experiment gap and provide a framework for predicting microbial responses to environmental stress, which might have potential implications for phage therapy, microbiome management, and climate-impacted community resilience as well.
\end{abstract}

\maketitle

%======================================================================
%  MAIN TEXT
%======================================================================
\section{\label{sec:In}INTRODUCTION}
Bacteriophages (phages), the most abundant biological entities on Earth, are fundamental architects of the structure and function of microbial ecosystems~\cite{RN96,doi:10.1073/pnas.96.5.2192,pathogens8030100}. 
As obligate bacterial parasites, they exert immense selective pressure on their hosts through lytic infection cycles, directly modulating bacterial population densities, community composition, and genetic diversity, etc.~\cite{https://doi.org/10.1111/j.1461-0248.2011.01624.x}. 
This dynamic predator-prey relationship fuels a relentless co-evolutionary arms race: Bacteria deploy sophisticated defense mechanisms, including CRISPR-Cas adaptive immunity, restriction-modification systems, and abortive infection strategies, while phages counter-adapt via rapid genomic diversification (e.g., mutations in receptor binding proteins or acquisition of anti-CRISPR proteins)~\cite{RN98,RN99}. 
Collectively, these interactions are pivotal drivers of microbial evolution, bio-geochemical cycling, and the emergence of bacterial pathogenicity, etc~\cite{RN97}.

A key defensive strategy involves bacteria modifying cell surface receptors to prevent phage adsorption. 
Initial phage infection requires the specific binding of viral adhesion proteins to bacterial surface receptors, which are frequently essential for core physiological processes such as nutrient uptake~\cite{RN102}. 
By altering receptor structure or abundance, bacteria can evade recognition by phages, but this often imposes a fitness trade-off. 
Since many receptors are critical for metabolism and viability, their loss or modification can impede growth and reduce competitive fitness, particularly under resource-limiting conditions~\cite{Yang2025-MicrobiologySpectrum-adaptivecostsofphageresistance,Chen2024-ISME-Trade-offs-between-receptor-modification-and-fitness,RN103}.

Beyond passive growth reduction driven by fitness costs in defensive trade-offs, variations in bacterial growth rates themselves directly modulate phage resistance~\cite{v13040656}. 
Evidence indicates an inverse relationship between bacterial growth rate and phage survival: rapidly growing populations exhibit significantly reduced survival rates after phage exposure~\cite{RN107,RN108}, while slowly growing cells show enhanced resilience under specific conditions~\cite{RN105}. 
Complementing this phenomenon, bacterial persistence, a transient dormant phenotypic state characterized by growth arrest, serves as a distinct adaptive strategy. 
Persister cells, a specialized subpopulation tolerant to phage-induced lysis, enable population survival by evading phage replication cycles through metabolic dormancy~\cite{RN106}.

In natural environments, abiotic factors such as temperature, pH, and nutrient availability exhibit considerable spatiotemporal variability, profoundly influencing phage-bacteria interactions. 
For instance, changes in temperature, salinity, pH, and organic matter content have been shown to affect key steps of phage infection, including lytic and adsorption activities in various settings~\cite{RN109,RN110}. 
Critically, nutrient availability directly modulates bacterial growth rates, which in turn shape the outcomes of phage infection~\cite{RN111}. 
In addition, periodic environmental variations, such as daily cycles in nutrient levels or temperature, can drive the synchronization between predator and prey populations. 
This temporal alignment, which involves the entrainment of population dynamics to external forcing, can either amplify or dampen the amplitudes of oscillations, thereby affecting the stability of coexistence and subsequent evolutionary outcomes~\cite{RN112,RN113}.

Mathematical modeling, particularly through systems of ordinary differential equations (ODEs), has proven instrumental in formalizing, simulating, and predicting the nonlinear dynamics inherent in phage-bacteria interactions.
A rich body of theoretical work employing diverse ODE models has elucidated complex system behaviors, including fixed points, (quasi-)periodic oscillations, and chaotic dynamics~\cite{eb7795aa-0466-3889-b47a-8725c2104293,BOLDIN2022JoTB,doi:10.1073/pnas.2414229121}.
These models are instructive in identifying critical parameters that govern interaction outcomes, such as burst size and latent period~\cite{10.3389/fmicb.2021.637490}. 
Furthermore, some extended frameworks taking into account phenotypic heterogeneity and spatial structure have provided deeper insights into the mechanisms that sustain diversity and facilitate evolutionary adaptation~\cite{RN117,RN116}.

However, most existing theoretical frameworks predominantly assume closed and stable environments or simulate dynamics within controlled chemostats, in stark contrast to natural microbial ecosystems, which usually experience continuous fluctuations in nutrient availability, temperature, pH, and other abiotic factors. 
Such environmental variability can displace populations from idealized equilibrium states, and may fundamentally alter their ecological and evolutionary trajectories, thereby challenging the predictions derived from traditional steady-state analyses.

Moreover, while theoretical studies have productively emphasized central phage traits (e.g., infection strategy, latency period, and lysis-lysogeny probability, as key determinants of viral fitness~\cite{BRUCE20215046,BOLDIN2022JoTB,10.1093/pnasnexus/pgad431,10.3389/fmicb.2017.01386}), they often implicitly treat bacterial hosts as static or passive resources. 
This perspective overlooks the dynamic influence of intrinsic bacterial characteristics, including growth physiology, metabolic costs of defense expression, and phenotypic plasticity, all of which actively affect the phage-bacteria interaction dynamics~\cite{CASTERS2024102481}. 
Particularly, under fluctuating environmental conditions, these bacterial traits not only respond to changes, but also substantially shape co-evolutionary outcomes~\cite{Nguyen2021-NatureCommunications-bacterial-growth-nutrient-fluctuations,Schwartz2023-ISME-coevolution-seed-bank}.

To bridge these gaps, we develop in this work a minimal ODE-based framework that explicitly integrates key bacterial traits and periodic environmental forcing into the phage-bacteria interaction dynamics.  
By doing so, we seek to answer three critical questions: 
(1) How do intrinsic bacterial traits govern short-term population dynamics and long-term coexistence? 
(2) How does periodic environmental forcing alter these outcomes and through which types of mechanistic pathways? 
(3) How do bacterial traits modulate the system's sensitivity to external environmental fluctuations? 
By resolving these questions, we expect to establish an efficient theoretical framework to investigate the intricate phage-bacteria dynamics with and without environmental forcing, highlighting the active and adaptive role of the bacterial hosts.

\section{\label{sec:Me} Model and Simulation Method}
\subsection{Model}
\begin{figure}
\includegraphics[width=8cm]{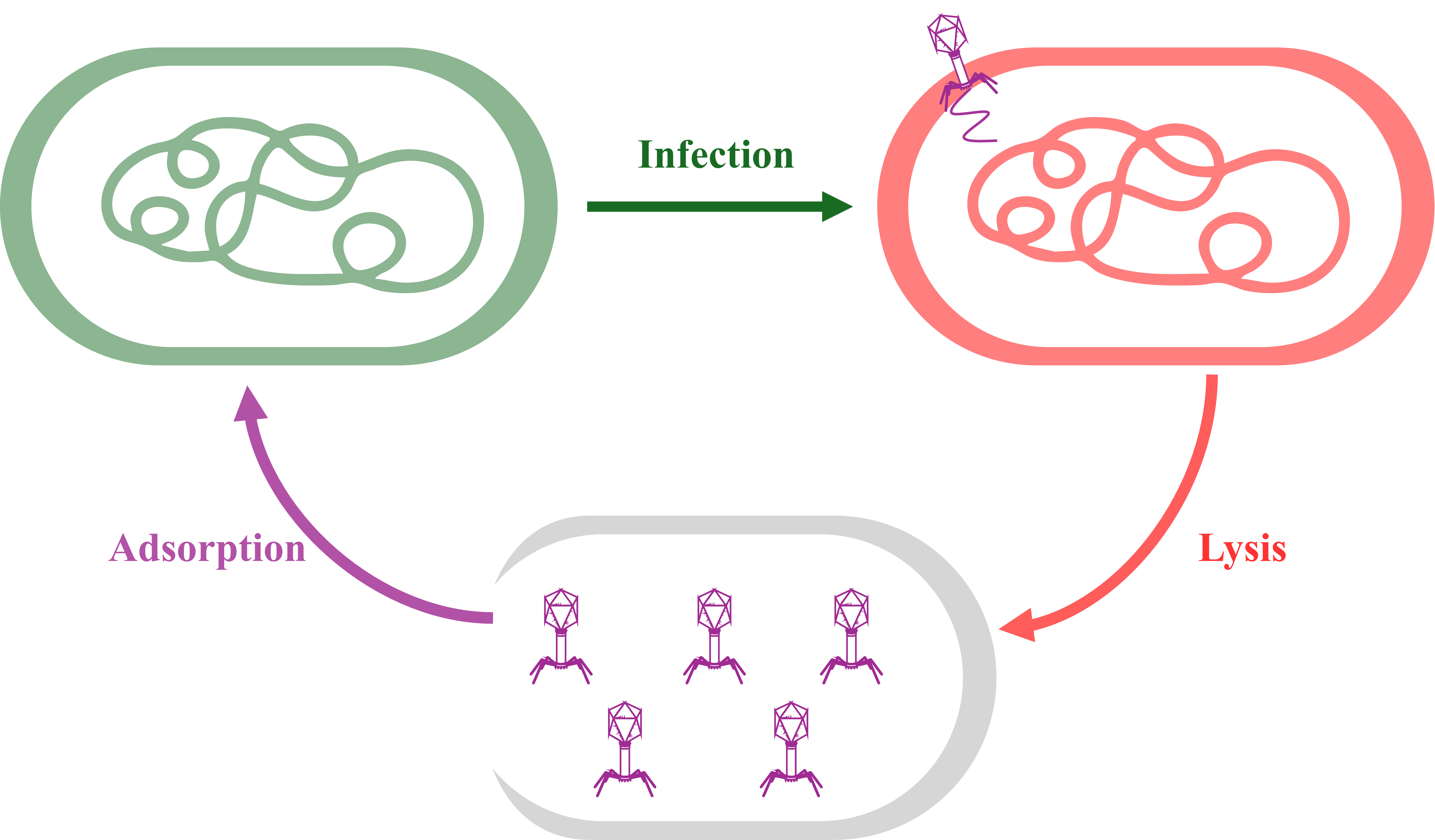}
\caption{\label{fig:mod1}
Schematic illustration of the lytic phage infection cycle. 
(1) Free phage particles are adsorbed onto the surface of uninfected bacterial host $B$ (top left). 
(2) Successful adsorption converts $B$ into infected intermediates $I$ (top right). 
(3) Maturation of $I$ culminates in cell lysis, releasing progeny virions that re-initiate infection (bottom).}
\end{figure}

To investigate the interplay between bacterial traits (we here mainly account for the two most important ones: adsorption efficiency and growth rate) and environmental fluctuations in shaping phage-bacteria competition dynamics, we first construct a deterministic ODE framework. 
In our current study, we focus exclusively on lytic phages, which propagate by adsorbing onto susceptible bacteria, hijacking intracellular machinery for replication, and ultimately lysing infected hosts to release progeny virions (illustrated schematically in Fig.~\ref{fig:mod1}). 
The bacterial population \(B\) grows logistically with a rate $r$ under an environmental carrying capacity \(K\), while phages are adsorbed onto bacteria with a rate $a$, converting uninfected cells into infected intermediates \(I\).

After a latent period, the infected bacteria undergo lysis with a rate \(\alpha\), and will produce \(\Omega\) newly assembled phage particles per lysed cell.
The existence of the intermediate state \(I\) characterizes the latent period of phages, and the rate of departure from \(I\) reflects the duration of this latent period.
The amount of free phages (\(P\)) increases through burst-releasing from lysed cells, and decreases through adsorption to new hosts or via natural decay at a rate \(\delta\). 
Taken together, the phage-bacteria dynamics is governed by the following equations:
\begin{equation}
\label{eq:model} 
    \begin{aligned}
    \frac{\mathrm{d}B}{\mathrm{d}t} &= \underbrace{r \left(1 - \frac{N}{K}\right) B}_{\text{Growth} }  - \underbrace{aBP}_{\text{Infection} } \\
    \frac{\mathrm{d}I}{\mathrm{d}t} &= \underbrace{aBP}_{\text{Infection} } - \underbrace{\alpha I}_{\text{Lysis} } \\
    \frac{\mathrm{d}P}{\mathrm{d}t} &= \underbrace{\Omega \alpha I}_{\text{Lysis} } - \underbrace{aBP}_{\text{Adsorption} } - \underbrace{\delta P}_{\text{Decay} },
    \end{aligned}
\end{equation}
where \(N=B+I\) denotes the total number of bacterial populations, including both susceptible and infected (not yet lysed) ones.
Bacteria can evolve surface receptor modifications to resist phage adsorption, a defense mechanism typically accompanied by an associated growth cost. 
We examine explicitly how two bacterial traits, specifically the adsorption rate (\(a\))  and the intrinsic growth rate (\(r\)), influence the competitive dynamics. 
Furthermore, we also take into account the effect of environmental fluctuations, which are typically incorporated by rendering the carrying capacity \(K\) as a time-dependent function \(K(t) = K_0 + AK_0 \sin(2\pi ft)\), introducing periodic fluctuations superimposed on the baseline capacity \(K_0\). 
Here, the amplitude \(A\) quantifies the relative intensity of environmental fluctuations, while the frequency \(f\) governs the oscillation rate.

\subsection{Simulation details}
For convenience, we consider in Eq.~(\ref{eq:model}) the rescaled variables: $\hat{K} = K / K_r$, $\hat{B} = B/ K_r$, $\hat{I} = I / K_r$, $\hat{P} = P / K_r$, and $\hat{a} = a K_r$, which are just structurally identical equations as the original ones. 
The parameter values used in our studies are mainly adopted from Ref.~\cite{2025GoelVirusEvolution}, and their biological interpretations and numerical settings are summarized in Table~\ref{tab:para}.
Unless otherwise specified, our presented simulations were performed in MATLAB R2022a using the built-in ODE solver.  
To ensure biological significance, an extinction threshold $\epsilon$ is imposed on each type of species such that their population densities decreasing to less than 1 individual per unit volume are truncated to zero. 
MATLAB codes that support the findings of this article are openly available at~\cite{Luo_code}.

\begin{table*}
\caption{\label{tab:para}%
Model parameters.
}
\begin{ruledtabular}
\begin{tabular}{llll}
\textrm{Parameter}&
\textrm{Description}&
\textrm{Value}&
\textrm{Units}\\
\colrule
r & Bacterial intrinsic growth rate & 0--2 & $\mathrm{hr}^{-1}$ \\
K & Environmental carrying capacity & 0--$2\times 10^8$ & $\mathrm{cells}\cdot \mathrm{mL}^{-1}$ \\
a & Adsorption rate & $10^{-12}$--$10^{-9}$& $\mathrm{mL}\cdot \mathrm{phage}^{-1}\cdot \mathrm{hr}^{-1}$\\
$\alpha$ & Cell lysis rate & 2 & $\mathrm{hr}^{-1}$ \\
$\Omega$ & Burst size  & 50 & $\mathrm{phages}\cdot \mathrm{cell}^{-1}$ \\
$\delta$ & Phage decay rate & 0.04 & $\mathrm{hr}^{-1}$ \\
$\mathrm{K_r}$ & Reference carrying capacity & $ 10^8$ & $\mathrm{cells}\cdot \mathrm{mL}^{-1}$ \\
$\hat{\mathrm{K}}=\mathrm{K}/\mathrm{K_r}$ & Rescaled carrying capacity & 0--2 & -- \\
$\hat{a}=a\cdot \mathrm{K_r}$ & 	Effective adsorption rate & $10^{-4}$--$10^{-1}$ & $\mathrm{hr}^{-1}$ \\
\end{tabular}
\end{ruledtabular}
\end{table*}

\section{Results and Analysis}

\subsection{Host Traits, Transient Dynamics, and Extinction Thresholds Control Coexistence and Collapse}

\begin{figure*}
\includegraphics[width=16cm]{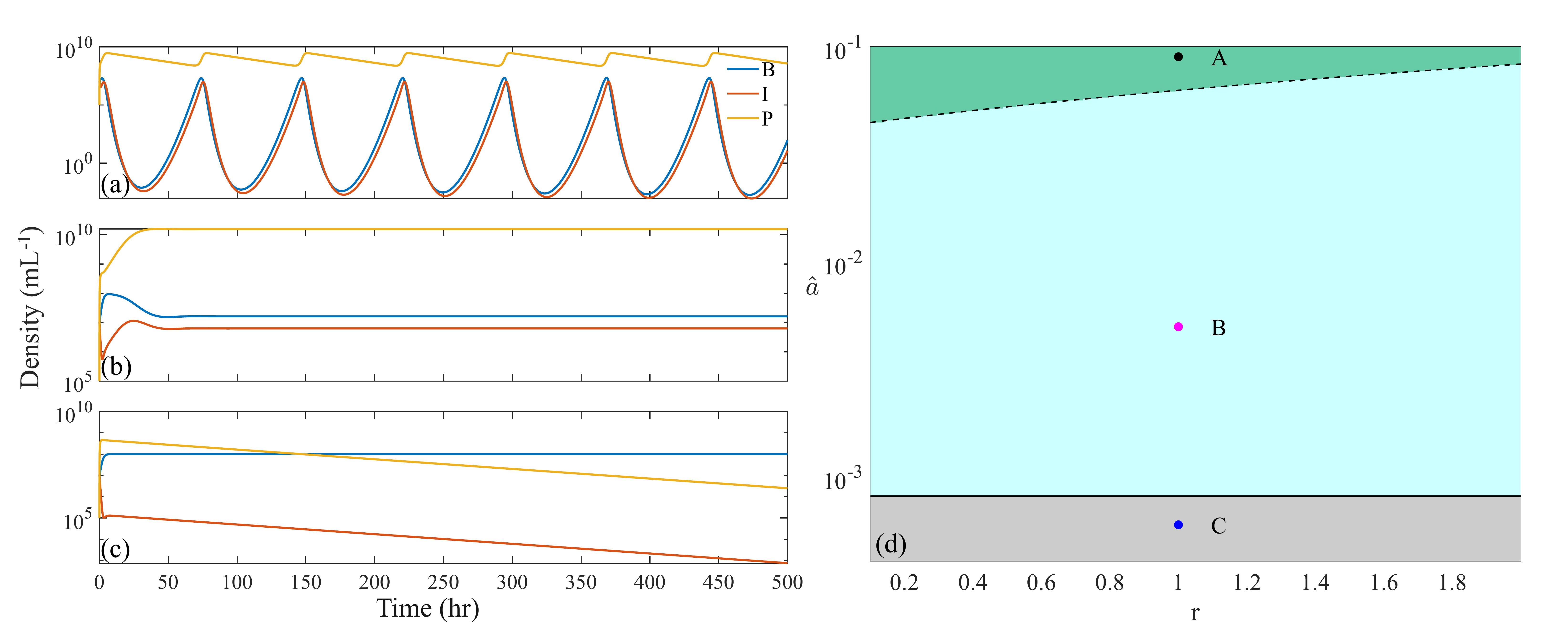}
\caption{\label{fig:RE1}
Dynamics and stability of the bacteria-phage system in the absence of an extinction threshold. 
(a)--(c) Time series for three different phage adsorption rates: (a) \(\hat{a} = 0.09\), (b) \(\hat{a} = 0.005\), and (c) \(\hat{a} = 0.0006\). 
The dynamics of the bacterial population (B, blue), infected bacteria (I, red), and phage (P, yellow) exhibit qualitatively distinct behaviors: a stable limit cycle (a), stable coexistence at equilibrium (b), and bacterial dominance with phage extinction (c). 
(d) Phase diagram in the \(r-\hat{a}\) parameter space, illustrating the regions of different dynamical behavior. 
The solid line represents the analytical threshold \(a_{\text{crit}} = \delta / [K (\Omega - 1)]\), below which phage invasion is impossible. 
The dashed line indicates the numerically determined boundary where the leading eigenvalue of the coexistence equilibrium acquires a positive real part, indicating the arise of a Hopf bifurcation. 
Gray region: bacteria persist, but with phage extinction; 
cyan region: stable coexistence of bacteria and phages; 
green region: limit cycle oscillations of the two species. 
Points A, B, and C denote the parameter sets inspected in panels (a), (b), and (c), respectively. 
Parameters: \(\hat{K} = 1\), initial conditions \(B(0) = 10^7\), \(I(0) = 10^7\), \(P(0) = 10^5\), and all other parameters as listed in Table~\ref{tab:para}.}
\end{figure*}

The asymptotic dynamics of bacteria--phage systems in constant environments have been extensively studied in classical ecological theory. 
Consistent with previously established frameworks\cite{Doekes2021eLife,BOLDIN2022JoTB,Stewart1984TPB,2019WahlEvo}, our linear stability analysis of the continuous deterministic limit reveals three well-known dynamical regimes in the $r$--$a$ parameter space (Fig.~\ref{fig:RE1}, also see the SM for detailed derivations). 
For low adsorption rate $a$, phages cannot successfully invade the host population, resulting in bacterial dominance and phage extinction (Fig.~\ref{fig:RE1}~(c)). 
As $a$ increases beyond the analytical threshold $a_{\text{crit}}$ for phage invasion, the system shifts to a regime of stable coexistence (Fig.~\ref{fig:RE1}~(b)). 
Further changing of $r$ or $a$ drives the coexistence equilibrium through a Hopf bifurcation, producing sustained limit-cycle oscillations (Fig.~\ref{fig:RE1}~(a)).
Although the linear stability analysis framework correctly predicts the long-term asymptotic attractors, it relies on the assumption of continuous population changes and ignores the demographic stochasticity that exist in finite microbial communities.

In real ecological environments, populations cannot recover from extremely low densities. 
To account for this fact, we consider a finite extinction threshold effect: if the density of any population falls below $\epsilon$, it is irreversibly lost. 
This biologically motivated modification fundamentally determines the system's fate, generating a bistable regime even within parameter regions that are linearly predicted to support stable coexistence (Fig.~S7). 
%the final outcome is closely affected by the transient dynamics during the initial stage, since bacterial population densities undergo sharp temporary dips during the phage outbreak. 
%If these transients breach $\epsilon$, the system is permanently driven to extinction; if they remain above it, trajectories converge to persistent coexistence or oscillations. 
Basically, we find that the long-term state depends sensitively on the initial densities of susceptible ($B_0$) and infected ($I_0$) bacteria, but is insensitive to the initial phage load ($P_0$), owing to rapid phage amplification (burst size $\Omega$). 
The complete structure of the basins of attraction, which explicitly quantifies the extinction-prone region embedded within the linearly stable regime, is provided in Fig.~S1 in the Supplementary Material. 
This threshold-dependent bistability and initial-condition-driven survival probability represent a key departure from classical continuous models.

\begin{figure*}[htp]
\includegraphics[width=16cm]{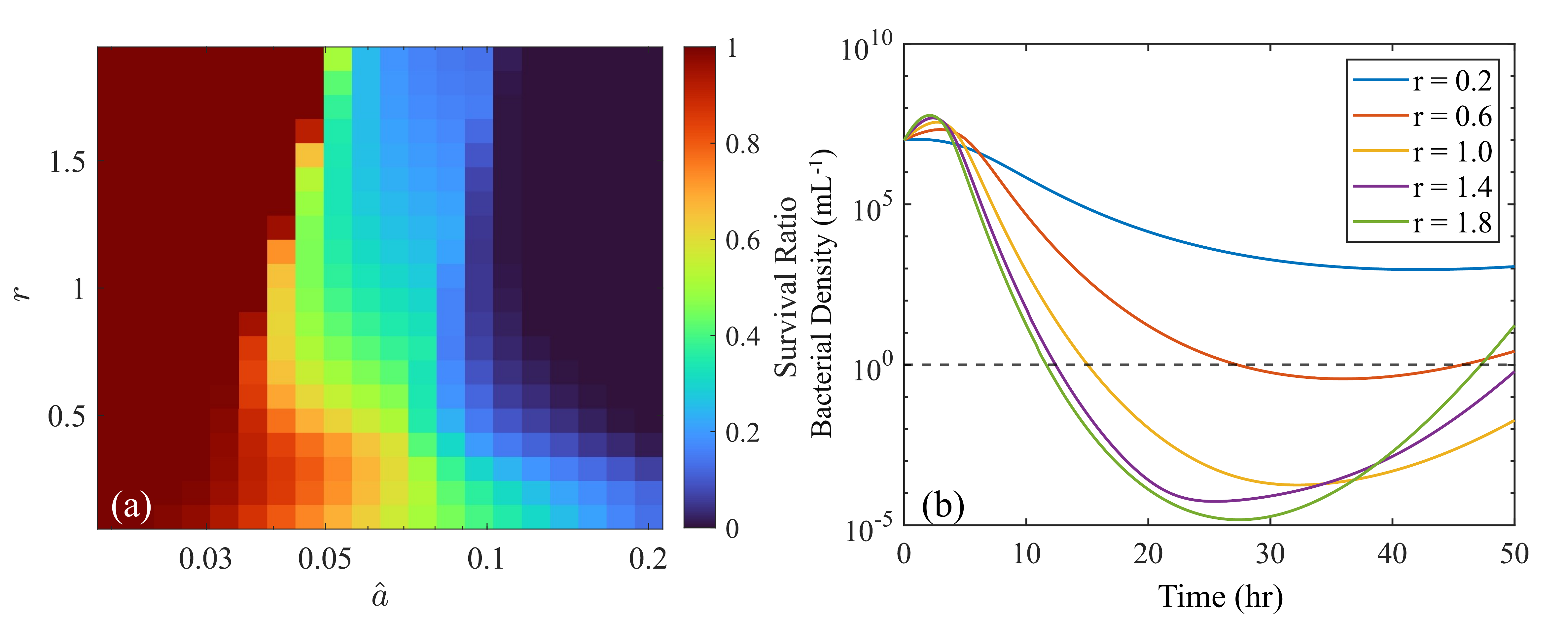}
\caption{\label{fig:RE3}
Dependence of the survival phase on the bacterial growth rate $r$ and phage adsorption rate $a$. 
(a) Heatmap of the survival probability for bacteria-phage coexistence. 
Color intensity represents long-term survival probability for different combinations of the parameters $r$ and $a$. 
High growth rates maximize the survival chance at low phage pressure (top-left), while low growth rates become favorable under high phage pressure (bottom-right), revealing a competitive reversal in defense strategies. 
(b) Representative time series of the simulations for the bacterial densities without an extinction threshold at a fixed adsorption rate \(\hat{a} = 0.07\). The black dashed line indicates the extinction threshold \(\epsilon = 1\). }
\end{figure*}

First, we quantify how host life-history traits shape the population survival chance under predation pressure. 
By systematically computing survival probabilities across different combinations of $B_0$--$I_0$, we uncover a competitive reversal in optimal defense strategies: under low phage pressure, a high growth rate $r$ maximizes bacterial persistence, whereas under high adsorption rates, slowly-growing strains exhibit significantly higher long-term survival probabilities (Fig.~\ref{fig:RE3}~(a)).
%This result arises from the coupling between the growth dynamics and the extinction threshold. 
This result can be reasonably understood as follows.
Fast growth rate generally accelerates the population recovery, but inevitably amplifies the magnitude of transient oscillations, which very likely results in sharp crashes of the population densities to a value below $\epsilon$, triggering system-wide collapse. 
In contrast, slower growth naturally dampens these transients, keeping populations safely above the extinction threshold and enabling stable coexistence, even at lower overall abundances (Fig.~\ref{fig:RE3}~(b)).

%Together, these results bridge theoretical dynamics and empirical observations in phage-bacteria ecology. 
Thus, by explicitly incorporating a finite extinction threshold, we shows that a reduced bacterial growth rate is not merely a metabolic cost.
Rather, it acts as an active dynamical buffer that mitigates extinction risk under intense predation pressure. 
By reducing transient oscillation amplitudes, a slower growth strategy effectively stabilizes the community against threshold-driven collapse. 
This provides a mechanistic explanation for the widely observed trade-offs between rapid proliferation and phage resistance~\cite{RN107,RN108,RN106,RN105}, showing how host life-history traits can greatly change microbial community persistence. 
This finding is analogous to the celebrated ``law of the weakest" observed in cyclic competition models, where the species with the lowest reproduction--predation rate exhibits the highest fixation probability in finite populations~\cite{Berr2009PRL,West2018PRE,Frean2001PoRSB,Reichenbach2006PRE}. However, we emphasis that our findings stem from a different dynamical mechanism. 
In cyclic models, survival is governed by demographic noise driving trajectories toward absorbing boundaries along neutrally stable orbits. 
By contrast, extinction in our host--phage system is just dictated by deterministic dynamics with a natural population amount restraint.

\subsection{Environmental Fluctuations Alter System Dynamics}
Having figured out the system's dynamical behaviors under constant conditions, we next investigate a more ecologically relevant scenario by introducing sinusoidal variation in the environment carrying capacity, a key parameter that influences the bacterial growth. 
This is conveniently implemented by setting \(K(t) = A K_0 \sin(2\pi f t) + K_0\), where \(A\) and \(f\) represent the amplitude and frequency of the environmental forcing, respectively.

\begin{figure*}
\includegraphics[width=15cm]{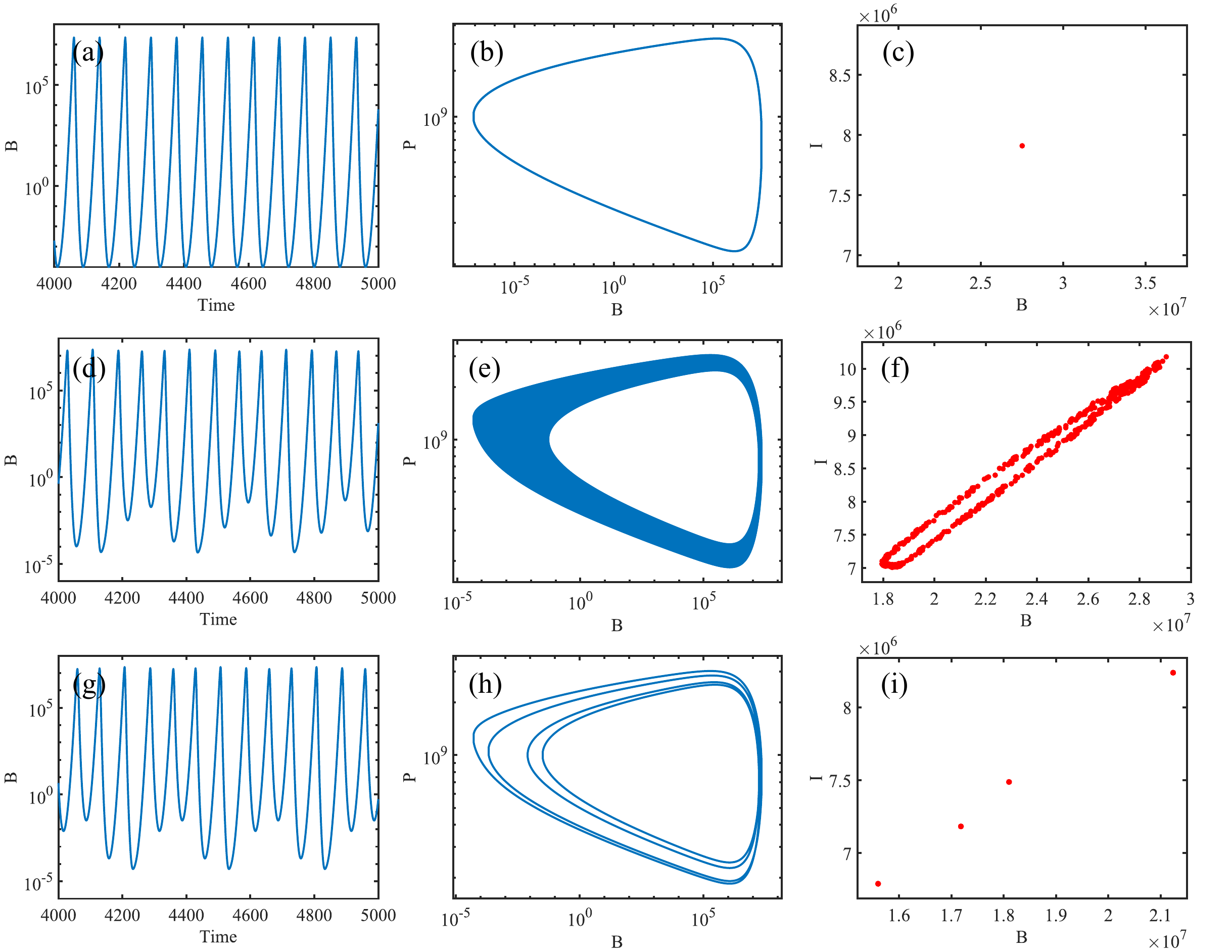}
\caption{\label{fig:RE4-2}
Response of the phage-bacteria system under periodic environmental forcing with different amplitude. 
Left column: Typical time series of bacteria \(B\); Middle column: Phase portraits in the \(B\)--\(P\) plane; 
Right column: Poincar\'e section diagram of the phase trajectories obtained by fixing \(P\). 
Upper row (\(A = 0\)): The system exhibits a stable limit cycle with periodic oscillation (a), forming a closed orbit in the phase plane (b), and lefting a single point in the Poincar\'e section (c). 
Middle row (\(A = 0.35\)): Quasi-periodic dynamics is observed (d), showing oscillations that do not repeat exactly; the phase portrait forms a densely filled torus (e), and the points in the Poincar\'e section forms a closed curve (f). 
Bottom row (\(A = 0.4\)): A multi-periodic state (here, a 4-period) emerges (g), appearing as separate closed orbits in the phase plane (h), and four discrete points are observed in the Poincar\'e section (i). 
Parameters: \(f = 0.01\), \(r = 1\), \(\hat{K_0} = 1\), \(\hat{a} = 0.1\).}
\end{figure*}

Our simulations reveal that environmental fluctuations have profoundly different impacts depending on the distinct steady states of the deterministic system (i.e., without fluctuations \(A=0\)). 
For parameter regimes where the deterministic system converges to stable fixed points, the introduction of moderate fluctuations of the environment just perturbs the system slightly, resulting in small-amplitude, regular oscillations around the former equilibrium points (Fig.~S8). 
By contrast, for systems operating in the limit cycle regime (deterministic oscillations, c.f., Fig.~\ref{fig:RE1}~(a)), periodic environmental forcing significantly alters the long-term dynamical behavior. 
We obtain versatile stationary dynamics of the phage-bacteria system, which is critically dependent on the coupling between the intrinsic oscillation frequency \(f_c\) of the limit cycle solution and the external forcing frequency \(f\) of the environment.
Specifically, we observe the emergence of multi-periodic states and also the quasi-periodic dynamics of the phase trajectories of the system in the stationary state; see Fig.~\ref{fig:RE4-2}. 
The different outcomes are verified by the diverse diagrams of points left in the Poincar\'e section, which is constructed by intersecting phase trajectories with the plane \( P = \text{const.} \). In simulations, this constant is taken as the average value of $P$ over a period of steady state.
And only those intersections satisfying the unidirectional crossing condition \( \dot{P} > 0 \) are retained, which guarantees a well-defined return map. 
Meanwhile, initial transients are discarded to ensure convergence to the stationary state.

\begin{figure}
\includegraphics[width=7.7cm]{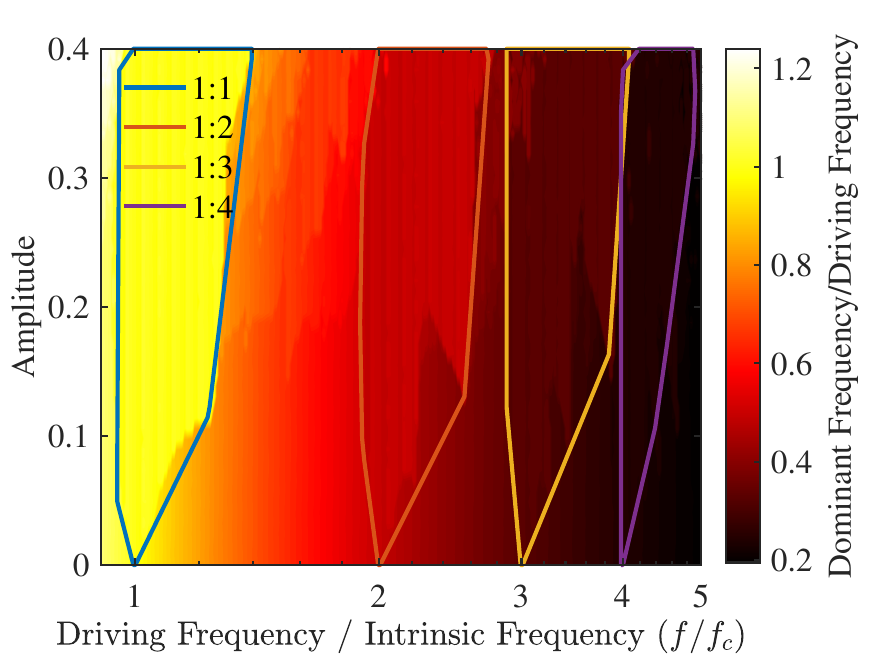}
\caption{\label{fig:RE4-3}
Frequency locking and the emergence of Arnold tongues under periodic environmental forcing. 
The heat map illustrates the dominant oscillation frequency of the bacterial population (expressed as a ratio to the driving frequency) across different driving frequencies (x-axis, normalized by the intrinsic frequency \(f_c\)) and forcing amplitudes (y-axis). 
The blue line indicates the convex hull of the region where the system's dominant frequency matches the driving frequency with a 1:1 ratio (relative error $\leq$ 0.5\%), representing the primary frequency locking. 
In addition, convex hulls in other colors mark the regions of higher-order harmonic locking: orange for 1:2, yellow for 1:3, and purple for 1:4 frequency ratios, respectively. 
The primary locked region expands as the forcing amplitude increases, forming a characteristic Arnold tongue, which signifies the occurrence of synchronization between the system and the external periodic forcing.
Parameters: \(r = 1\), \(\hat{K_0} = 1\), \(\hat{a} = 0.09\).}
\end{figure}

Theoretically, the unforced limit-cycle solution can be conceptualized as a self-sustained nonlinear oscillator. 
The periodic modulation of \(K(t)\) acts as an external driver, and the two subjects become coupled through the growth term \((1 - N/K(t))\) in the model, effectively forming a unidirectionally coupled oscillator system, which exhibits complex synchronization phenomena. 
As illustrated in Fig.~\ref{fig:RE4-3}, analysis of the forced system shows that the dominant oscillation frequency shifts away from its intrinsic value \(f_c\) by tuning the external forcing frequency.
Within specific regions of the forcing parameter space (amplitude \(A\) and frequency \(f\)), the system's frequency locks to the driving frequency (\(f\)), achieving an 1:1 frequency locking. 
The range of driving frequencies over which this locking occurs increases with the forcing amplitude \(A\), forming the characteristic Arnold tongue synonymous with synchronized oscillatory systems.

\subsection{Environmental Forcing can Mitigate Extinction via Resonance}\label{subs:RE5}
Based on the characterized dynamical regimes in Fig.~\ref{fig:RE1}~(d), we can now directly assess how periodic environmental forcing affects the ultimate fate of the bacterial population--specifically, whether it tends to drive the system towards extinction or sustain it.
To be specific, we quantify the risk of extinction by scanning the forcing parameters and recording the local minima of the time series of the bacterial population, $B_{\mathrm{min}}$, during the relaxation stage. 
A value of $B_{\mathrm{min}}$ above the threshold $\epsilon$ indicates sustained survival of the bacteria, while a value below it means extinction.

\begin{figure}[htp]
\includegraphics[width=7.7cm]{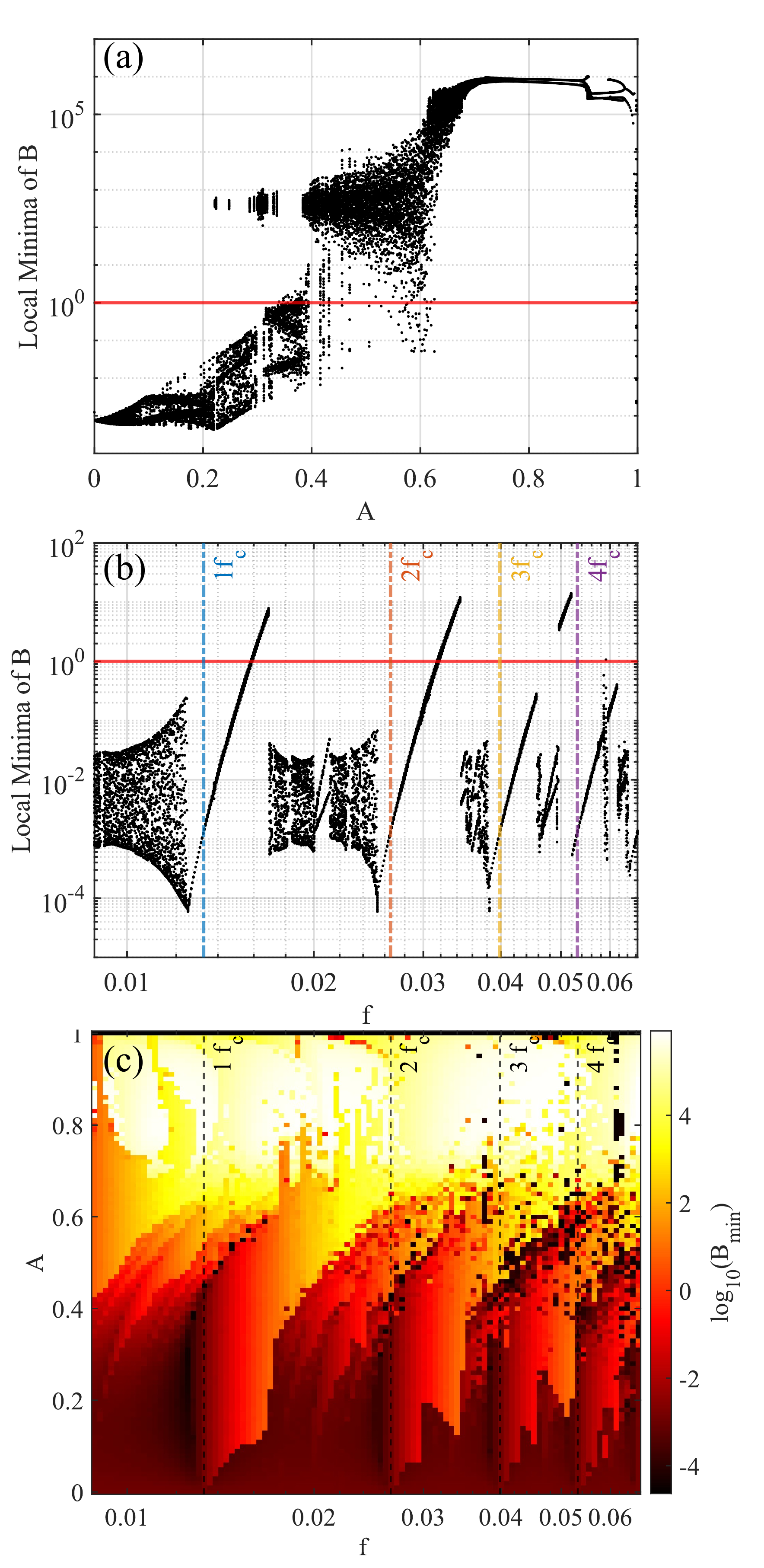}
\caption{\label{fig:RE5}
Environmental forcing modulates the risk of extinction of the bacteria through resonance.
(a) Local minima of the bacterial population $B_{\mathrm{min}}$ as a function of the external forcing amplitude ($A$) of the environmental capacity with a fixed frequency ($f = 0.02$). 
The red horizontal line indicates the extinction threshold ($\epsilon = 1$). 
(b) $B_{\mathrm{min}}$ as a function of the forcing frequency ($f$) at a fixed amplitude ($A$ = 0.1). 
The red horizontal line is the extinction threshold. 
Vertical dashed lines mark the integral multiples (1-4) of the system's intrinsic frequency ($f_c$). 
(c) Two-parameter phase diagram illustrates the dependence of the quantity $B_{\mathrm{min}}$ on different combinations of the forcing amplitude ($A$) and its corresponding frequency ($f$). 
The color scale represents the magnitude of $B_{\mathrm{min}}$. 
Vertical dashed lines correspond to the harmonics of $f_c$ as in the panel (b). 
The structure of the survival regions suggests that frequency locking to the environmental driver (as characterized by the Arnold tongue structure in Fig.~\ref{fig:RE4-3}) is the underlying mechanism that suppresses large oscillations of population densities and mitigates the risk of extinction.
Parameters: \(r = 1\), \(\hat{K_0} = 1\), \(\hat{a} = 0.09\).}
\end{figure}

Strikingly, the introduction of fluctuating environment not only merely perturbs the system, but also can fundamentally alter its evolving fate. 
Figure~\ref{fig:RE5}~(a) shows the results of an amplitude sweep at a fixed forcing frequency \(f\). 
For the unforced case (\(A=0\)), $B_{\mathrm{min}}$ is less than the threshold, confirming the extinction of bacteria. 
As the amplitude \(A\) increases from zero, the system initially remains in the extinction state, since $B_{\mathrm{min}}$ is still smaller than \(\epsilon\). 
However, a further increase in \(A\) leads to the presence of a counterintuitive rescue effect: the band of minima shifts upward, and beyond a certain critical amplitude, the entire band rises above the extinction threshold. 
This transition indicates the onset of frequency locking, where the external forcing tames the large, destructive oscillations of the native limit cycle. 
This resonant synchronization effect stabilizes the population density by preventing it from dipping below the critical level, thereby preventing the extinction of the bacterial population that is otherwise inevitable in the static environment. 
This rescue effect persists over a range of amplitudes of \(A\) until it becomes too large (\(A \to 1\)), where the carrying capacity \(K(t)\) itself causes a collapse and a sharp drop in $B_{\mathrm{min}}$ when it periodically approaches zero.

As shown in Fig.~\ref{fig:RE5}~(b), the influence of the forcing frequency \(f\) is equally profound, which unexpectedly induces non-monotonic survival responses of the bacterial population. 
For a low amplitude \(A=0.1\), extinction events (where $B_{\mathrm{min}}<\epsilon$) occur almost always over a wide range of frequencies $f$. 
However, distinct ``resonant windows" emerge when \(f\) is a rational multiple of the unforced limit cycle frequency \(f_c\), e.g., \(f = f_c\cdot n\), \(n \in \mathbb{Z}^+\). 
Within these windows, the onset of synchronization effectively reduces the amplitude of oscillation, narrowing the minima band of $B_{\mathrm{min}}$ and elevating it above \(\epsilon=1\). 
Consequently, the frequency-entrainment mechanism due to synchronization facilitates the stable coexistence of all the populations through stabilized periodic dynamics.

The two-parameter diagram in Fig.~\ref{fig:RE5}~(c) further summarizes these results, mapping the sustained bacterial population minimum across the \((A, f)\) parameter space. 
We find that the regions supporting stable survival, quantified by $B_{\mathrm{min}}>\epsilon$, correspond directly to Arnold tongue structures predicted by our model in Fig.~\ref{fig:RE4-3}. 
For a detailed comparison highlighting these structural correspondences, please refer to the Supplementary Material (Fig.~S9).
The characteristic geometry of these survival regions--narrowing to a point at low amplitudes and widening with increasing \(A\)--closely mirrors the classic signature of synchronization in driven nonlinear dynamical systems~\cite{1988FromClockstoChaos:TheRhythmsofLife}. 
Frequency locking within Arnold tongues suppresses large-amplitude oscillations, elevating the bacterial population minima and reducing its risk of extinction.
Thus, our results provide strong supporting evidence for the hypothesis that environmental variability can suppress inherent oscillatory instabilities and mitigate the risk of extinction~\cite{Peniston2020-ProceedingsoftheRoyalSocietyB-Environmental-fluctuations-promote-rescue,Stephanie2014-TrendsinEcologyEvolution-Evolutionary-rescue-changing-world}, but primarily within specific amplitude and frequency ranges that successfully promote a synchronized response.
From an evolutionary ecology perspective, this phenomenon may offer a mechanistic explanation for the observed synchronization between population amount and environmental fluctuations in nature~\cite{RN121,RN122,ijms18040873}--such dynamics could be interpreted as an adaptive outcome, where entrainment to external frequencies confers a survival advantage by preventing catastrophic population declines.

\subsection{Growth and Infection Traits Modulate the Response to Environmental Fluctuations}
The preceding analysis has established two key insights: First, that intrinsic bacterial traits (\(r\), \(a\)) govern dynamics and thereby determine the survival or extinction in a static environment; 
And second, that extrinsic environmental forcing can reshape long-term dynamics, either exacerbating or mitigating extinction risk through resonance and frequency locking. 
Thus, a critical and natural question arises: How do intrinsic bacterial traits modulate the system's response to extrinsic environmental fluctuations? 

\begin{figure*}
\includegraphics[width=16cm]{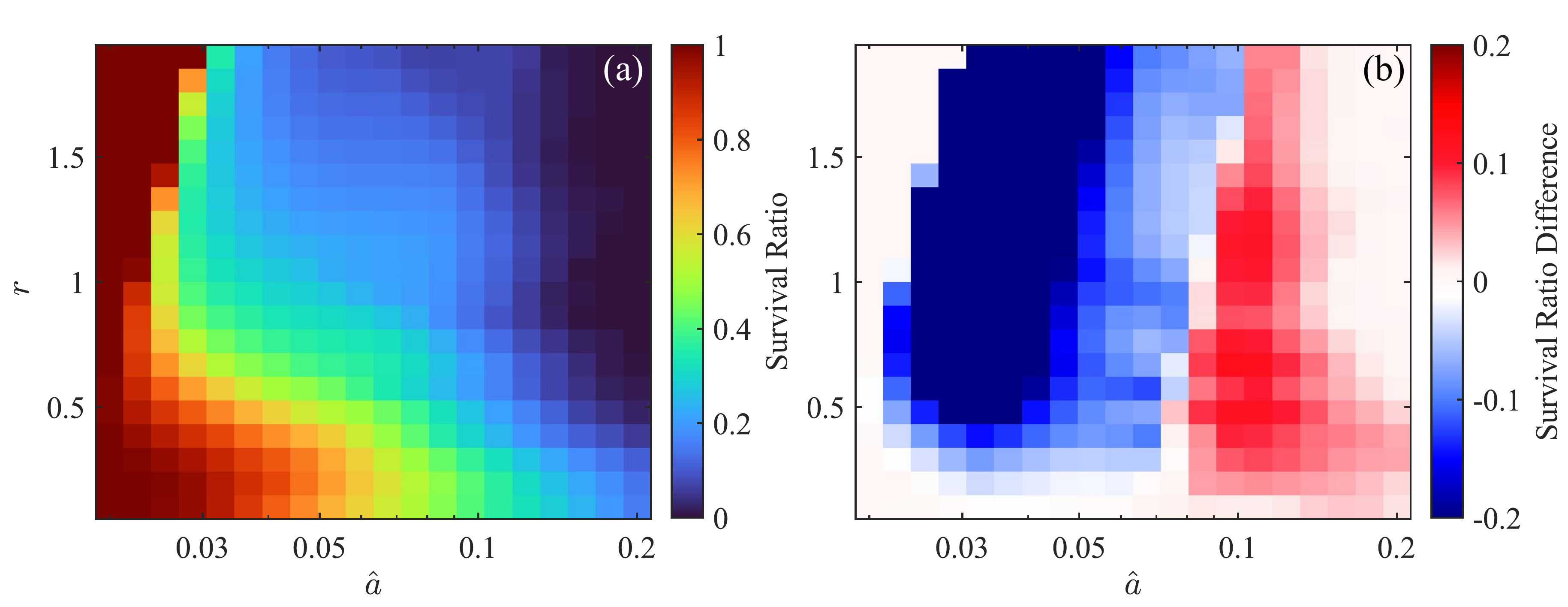}%  
\caption{\label{fig:RE6}
Impact of the environmental fluctuations on the bacterial survival. 
(a) Heatmap of the survival ratio under environmental forcing. 
(b) Heatmap of the survival ratio difference (\(\Delta S_{\mathrm{surv}} = S_{\mathrm{fluct.}} - S_{\mathrm{static}}\)) induced by the environmental fluctuations. 
The color scale is symmetrically clipped to \([-0.2, 0.2]\) for visual clarity; regions with \(\Delta S_{\mathrm{surv}} < -0.2\) are saturated in deep Blue. 
Blue (red) zones indicate the cases where fluctuations suppress (promote) the survival chance of bacteria, respectively. 
Parameters: baseline carrying capacity \(\hat{K}_0 = 1\), forcing amplitude \(A = 0.8\), frequency \(f = 0.05\), extinction threshold \(\epsilon = 1\).}
\end{figure*}

To address this problem, we employ the same basin-scanning methodology as in Fig.~\ref{fig:RE3}~(a) to detect the population dynamics, but now under significant sinusoidal forcing of the carrying capacity (\(A = 0.8\), \(f = 0.05\)).
The resulting fractions of successful survival of bacteria under different combinations of ($r$, $\hat{a}$) are summarized in Fig.~\ref{fig:RE6}~(a), which reveals a delicate interaction between traits and forcing. 
Overall, we find a decrease in the possibility of survival of the bacteria as $\hat{a}$ increases beyond a critical value, analogous to the scenario observed in the static environment.

However, unlike the case of static environment (Fig.~\ref{fig:RE3}), fluctuating environment induces a dual effect for the survival chance, as quantified in Fig.~\ref{fig:RE6}(b): 
(1) For low infection pressure (left side of the heatmap), environmental forcing reduces the probability of survival of bacteria, indicated by the Blue zones. 
(2) For high infection pressure (right side), environmental forcing, however, increases the probability of bacterial survival, evidenced by the red zones.

\begin{figure}
    \centering
    \includegraphics[width=7.7cm]{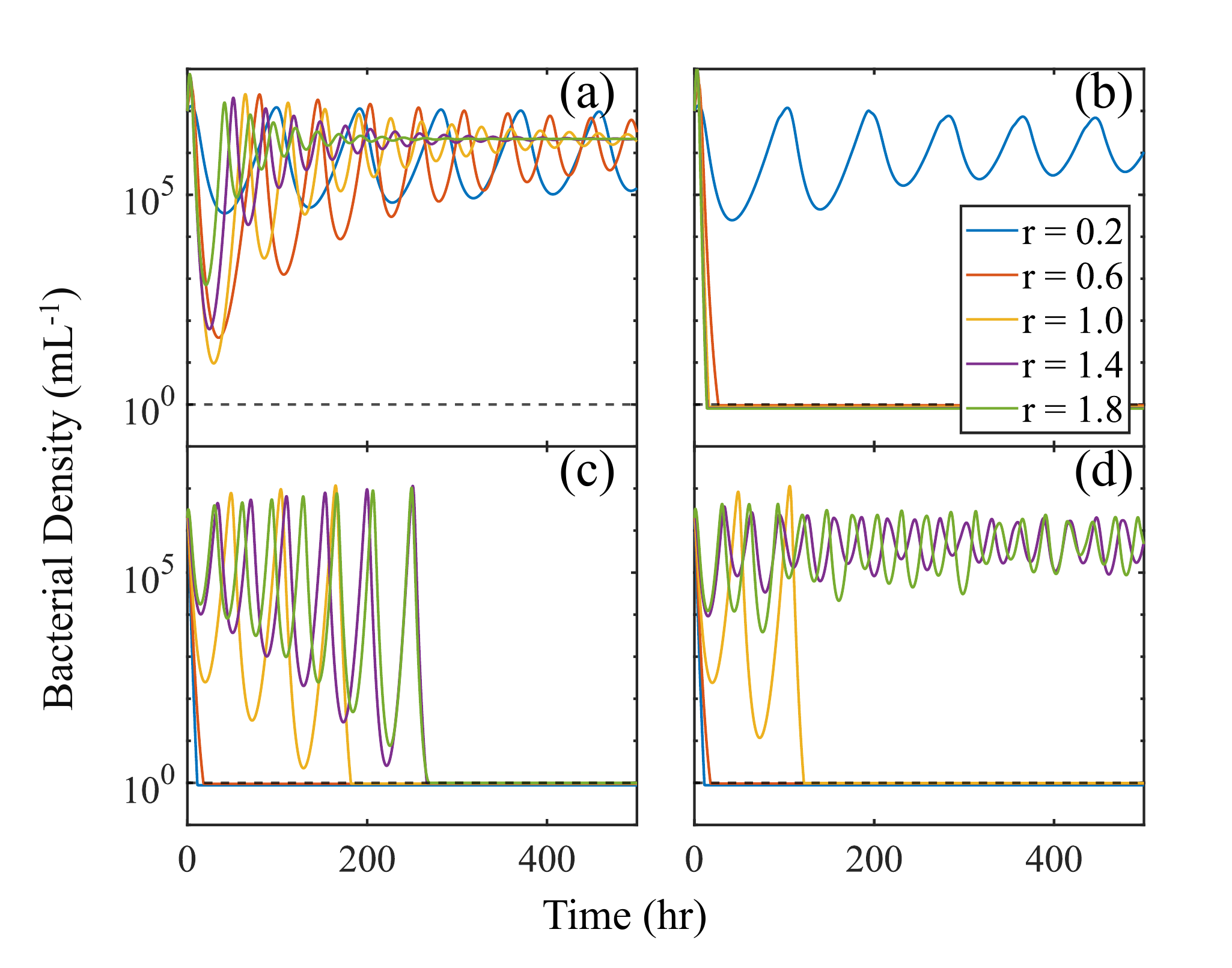}
    \caption{Typical time series of the bacterial population density illustrating how environmental fluctuations alter extinction dynamics. 
    (a)--(b) Low adsorption regime (\(\hat{a} = 0.038\)): 
    (a) In a static environment, bacteria maintain stable survival. 
    (b) Under fluctuating conditions, transient oscillations are amplified, driving the population below the extinction threshold. 
    (c)--(d) High adsorption regime (\(\hat{a} = 0.12\)): 
    (c) In a static environment, the population exhibits growing oscillations that inevitably lead to extinction. 
    (d) Environmental forcing suppresses these large-amplitude oscillations through frequency locking, stabilizing the density above \(\epsilon\) and demonstrating a clear rescue effect. 
    Parameters: \(\hat{K}_0 = 1\), \(A = 0.8\), \(f = 0.05\), \(\epsilon = 1\). 
    Initial conditions: (a)--(b) \(B_0 = 0.1\), \(I_0 = 0.05\), \(P_0 = 0.1\); 
    (c)--(d) \(B_0 = 0.02\), \(I_0 = 0.3\), \(P_0 = 0.1\).}
    \label{fig:RE6-2}
\end{figure}

To elucidate the mechanisms underlying these seemingly contrasting effects, we examine the time series of the bacterial density for different growth rates $r=0.2, 0.6, 1.0, 1.4, 1.8$, under both the conditions of high and low infection pressures.
The results are shown in Fig.~\ref{fig:RE6-2}.

Under low infection pressure, we find that environmental forcing can amplify the amplitude of initial transient oscillations of bacterial density, causing trajectories that would otherwise remain above the threshold to dip below it for almost all growth rates considered (except the smallest one $r=0.2$), thereby reducing the survival probability of bacteria, c.f., Figs.~\ref{fig:RE6-2}~(a) and (b). 
By contrast, under high infection pressure, we observe that environmental forcing is able to enhance the survival probability of bacteria by suppressing the sustained oscillation amplitudes of their densities via the frequency locking mechanism, see Figs.~\ref{fig:RE6-2}~(c) and (d). 
Comparing Figs.~\ref{fig:RE6}~(b) and (d), we find that the rescue effect, where environmental fluctuations are helpful to mitigate the risk of extinction, does not occur under the condition of low infection pressure, due to the absence of pronounced oscillatory dynamics of bacterial density.
Instead, it is the enlarged amplification of the amplitude of oscillation during the initial spreading stage of the phages that drives the bacterial population to extinction, see Fig.~\ref{fig:RE6}~(b).

We remark that the bacterial growth rate \(r\) serves as a critical modulator of these effects.
A phage-bacteria system with a sufficiently large \(r\), upper portion of Fig.~\ref{fig:RE6}, usually exhibits markedly enhanced sensitivity to environmental forcing. 
Specifically, for low-to-medium values of \(a\), large values of $r$ improve the transient amplification of detrimental oscillations, while for high adsorption rates \(a\), they instead favor a beneficial transition into a stabilized resonant state. 
From a mechanistic perspective, this arises because the growth term \(r(1 - N/K(t))\) scales the magnitude of the population response to oscillations in carrying capacity \(K(t)\) such that larger values of \(r\) would always produce proportionally greater dynamical responses to identical environmental perturbations.

Summing up, our results demonstrate explicitly that the impact of environmental fluctuations is not universal in the phage-bacteria interaction dynamics but rather is critically filtered by the intrinsic life-history traits of the bacterial hosts. Fluctuating environment can be either punitive or beneficial, acting as a double-edged sword.
On the one hand, it increases the risk of extinction for bacteria in the cases of low-to-medium infection pressure by destabilizing the transient dynamics of the two competitive species. 
On the other hand, it is counterintuitively capable of mitigating the extinction risk of bacteria under high infection pressure via facilitating resonant stabilization of the long-term dynamics. 
Interestingly, it is the growth rate \(r\) of the bacteria themselves that determines the sharpness of this sword such that faster-growing bacterial populations seem to experience the dualistic impact of environmental forcing more extremely.

\section{Conclusions and Discussions}
In summary, our minimal dynamic framework of the phage-bacteria interactions elucidates how intrinsic bacterial traits and environmental fluctuations jointly dictate the long-term ecological outcomes. 
In contrast to the work by Goel \textit{et al.}~\cite{2025GoelVirusEvolution}, which focused on temperate phage life-history evolution under binary feast-famine cycles, our study centers on obligately lytic phage-bacteria dynamics under continuous sinusoidal environmental forcing, with explicit focus on bacterial physiological traits as core modulators. 
By incorporating a biologically grounded extinction threshold, we move beyond conventional local stability analysis to reveal transient-driven bistability. Notably, we uncover a counterintuitive survival advantage of slower-growing bacteria under high infection pressure, as reduced growth dampens destructive population oscillations that would otherwise breach the extinction threshold, which is in line with the spirit of the celebrated ``law of the weakest'' observed in cyclic competition models~\cite{Berr2009PRL,West2018PRE,Frean2001PoRSB,Reichenbach2006PRE}.

Crucially, we show that environmental fluctuations are not merely perturbations but can fundamentally rewire system dynamics through the mechanism of resonance. 
Periodic forcing induces frequency locking (manifested as Arnold tongues), which effectively suppresses oscillation amplitudes and rescues bacterial populations from extinction. 
This resonance-mediated stabilization reveals a dual role of environmental variability: it amplifies extinction risk under low-to-moderate infection pressure but paradoxically promotes coexistence under high pressure. 
Such growth-rate-dependent modulation challenges the conventional view that environmental variability invariably destabilizes ecosystems, offering a mechanistic basis for the widespread phage--bacteria coexistence observed in nature.

Furthermore, we argue that the observed frequency locking is largely attributed to the delayed infection mechanism of the phage. 
Theoretically, it is the delay in the phage infection that drives the phage and bacterial populations to oscillate with certain intrinsic frequency. 
Whenever the frequency of environmental forcing becomes comparable to this intrinsic frequency, the frequency locking phenomenon naturally emerges, giving rise to non-trivial dynamical phenomena. 
We conjecture that such phenomena could be absent or significantly suppressed in a model assuming instantaneous infection and lysis.

We acknowledge that natural population dynamics are inherently subject to demographic stochasticity~\cite{McKane2005prl}. 
To account for this, we performed stochastic simulations using the Gillespie algorithm (see the Supplementary Material). 
Our preliminarily simulations under three distinct infection pressures yield dynamics qualitatively consistent with those predicted by the deterministic ODE model. 
Additionally, a rescue effect driven by the environmental fluctuations is observed. 
In contrast to the deterministic framework, extinction emerges naturally without requiring an extinction threshold, and the timing and frequency of population fluctuations exhibit intrinsic stochasticity. 
Here, we would like to remark that the effect of environmental fluctuations in the stochastic scenarios calls for a different theoretical characterization, as the deterministic concept of parametric resonance is no longer directly applicable.
%A systematic investigation of these stochastic features is reserved for future work.
Besides, natural environments rarely exhibit perfect periodicity~\cite{Silva2023pre, Pal2022pre}. 
Our supplementary tests with square-wave and noise-perturbed sinusoidal forcings reveal that the core dynamical regimes remain largely unchanged, except in very few special cases, thereby underscoring the robustness of the predictions from our  deterministic framework (please see the results summarized in the Supplementary Material). 
Thus, extending our current analysis to more realistic, multi-scale environmental spectra~\cite{Taitelbaum2020prl} remains a promising avenue for the future research.

Environmental fluctuations--such as daily cycles, thermal oscillations, and seasonal shifts--can effectively suppress the emergence of destructive oscillations via resonance-mediated rescue, thereby enabling stable coexistence. 
From an even broader perspective, the synchronized population dynamics frequently observed in microbial communities~\cite{RN121,RN122,ijms18040873} may not be mere stochastic artifacts, but rather adaptive responses shaped by these environmental rhythms. 
Introducing such periodic fluctuations into other systems with intrinsic oscillatory dynamics, such as the classic Lotka--Volterra predator--prey system, may yield more generalizable insights into how environmental variability governs stability in complex ecological systems.

Our current model focuses exclusively on lytic phages, excluding temperate and lysogenic phages, which are also widespread in the microbial world, and the involvement of them would definitely provide more insightful views on the phage-bacteria dynamics~\cite{Mancuso2021elife}. 
This issue is highly challenging, since the life history traits of the latter two types of phages are even more complicated than the lytic ones~\cite{Kim2017RPP, Kim2025pre}. 
Future work integrating bacterial diversity and multi-species interactions will further refine the theoretical predictions.

To conclude, we have established a meaningful theoretical framework to study the intricate phage-bacteria dynamics under environmental stress, highlighting the active and adaptive role of bacterial hosts in shaping their ecological dynamics. 
By uncovering how environmental fluctuations can counterintuitively stabilize phage-bacteria systems through resonance, we have provided a new perspective for understanding the dynamics of microbial population in natural ecosystems. 
These insights may ultimately inform strategies for managing microbial communities, with applications ranging from human health to environmental biotechnology.

\begin{acknowledgments}
We sincerely thank the reviewers for their insightful comments and constructive suggestions. 
Their valuable feedback has greatly improved the quality, logical rigor and overall presentation of this manuscript.

We acknowledge financial support from the National Natural Science Foundation of China (Grants No. 12375032, No. 12575039, and No. 12247101), and from the Fundamental Research Funds for the Central Universities (Grants No. lzujbky-2023-ey02, No. lzujbky-2024-11, and No. lzujbky-2025-jdzx07). 
This work was partly supported by the Longyuan-Youth-Talent Project of Gansu Province, by the Natural Science Foundation of Gansu Province (No.25JRRA799), and by the '111 Center' under Grant No. B20063. 
\end{acknowledgments}

\bibliography{bibliograph}

%======================================================================
%  SUPPLEMENTARY MATERIAL
%======================================================================
\clearpage
\onecolumngrid
\setcounter{figure}{0}
\setcounter{table}{0}
\setcounter{equation}{0}
\setcounter{page}{1}

\renewcommand{\thefigure}{S\arabic{figure}}
\renewcommand{\thetable}{S\arabic{table}}
\renewcommand{\theequation}{S\arabic{equation}}
\renewcommand{\thepage}{S\arabic{page}}
\renewcommand{\theHfigure}{S\arabic{figure}}
\renewcommand{\theHtable}{S\arabic{table}}
\renewcommand{\theHequation}{S\arabic{equation}}

\begin{center}
{\Large \textbf{Supplementary Material}} \\[5pt]
{\large \textbf{Frequency Locking to Environmental Forcing Suppresses Oscillatory Extinction in Phage-Bacteria Interactions}} \\[5pt]
Hao-Neng Luo, Zhi-Xi Wu$^*$, Jian-Yue Guan
\end{center}

\begin{quotation}
\noindent This Supplementary Material provides detailed mathematical derivations, numerical analyses, and stochastic validations supporting the findings on bacteria-phage interaction dynamics presented in the main text. 
First, we present a linear stability analysis of the three biologically relevant equilibrium points, deriving explicit conditions for local asymptotic stability and identifying the critical adsorption rate that governs phage invasion and Hopf bifurcations. 
Second, we introduce an extinction threshold to account for natural population size restriction, revealing how this realistic modification transforms the phase-space diagram and induces threshold-dependent bistability; we systematically map the geometry of basins of attraction and demonstrate their sensitivity to initial conditions and key parameters ($r$, $a$). 
Third, we validate the deterministic predictions using Gillespie stochastic simulations. 
Finally, we assess the robustness of our core conclusions under alternative environmental forcing scenarios, including Gaussian white-noise perturbations and non-sinusoidal (square-wave) periodic modulations, showing that the principal dynamical behaviors persist across diverse fluctuation profiles. 
Summarizing, these complemented analyses strengthen the theoretical foundation and ecological interpretability of the main results.
\end{quotation}

\section{Stability Analysis of Equilibrium Points}\label{app:Deri}

The dynamical system described by Eq.~(1) in the main text admits three biologically relevant equilibrium points. 
Their stabilities are rigorously analyzed through the classical linearization method and eigenvalue analysis of the Jacobian matrix evaluated at each equilibrium.

In particular, the system possesses the following equilibrium solutions:
(1) Extinction equilibrium: \( E_0 = (0, 0, 0) \)
(2) Bacteria-dominated equilibrium: \( E_1 = (K, 0, 0) \)
(3) Coexistence equilibrium: \( E_2 = (B^*, I^*, P^*) \), where \( B^*, I^*, P^* > 0 \) are positive solutions satisfying \( \dot{B} = \dot{I} = \dot{P} = 0 \).

The general Jacobian matrix of the system is given by:
\[
J(B, I, P) = 
\begin{bmatrix}
    J_{11} & J_{12} & J_{13} \\
    J_{21} & J_{22} & J_{23} \\
    J_{31} & J_{32} & J_{33}
\end{bmatrix},
\]
where the partial derivatives are explicitly:
\[
\begin{aligned}
    J_{11} &= r\left(1 - \frac{B + I}{K}\right) - \frac{r}{K}B - aP, \\
    J_{12} &= -\frac{r}{K}B, \\
    J_{13} &= -aB, \\
    J_{21} &= aP, \\
    J_{22} &= -\alpha, \\
    J_{23} &= aB, \\
    J_{31} &= -aP, \\
    J_{32} &= \Omega\alpha, \\
    J_{33} &= -aB - \delta.
\end{aligned}
\]

Evaluating the Jacobian at \( E_0 \) yields:
\[
J(0,0,0) = 
\begin{bmatrix}
    r & 0 & 0 \\
    0 & -\alpha & 0 \\
    0 & \Omega\alpha & -\delta
\end{bmatrix}.
\]
This upper triangular matrix has eigenvalues \( \lambda_1 = r \), \( \lambda_2 = -\alpha \), and \( \lambda_3 = -\delta \). 
Given \( r > 0 \), \( \alpha > 0 \), and \( \delta > 0 \) (standard parameter constraints), \( \lambda_1 > 0 \) while \( \lambda_2 < 0 \) and \( \lambda_3 < 0 \). 
The presence of a positive eigenvalue confirms that \( E_0 \) is a saddle point with a one-dimensional unstable manifold. 
Consequently, \( E_0 \) is unstable.

The Jacobian at \( E_1 \) is:
\[
J(K, 0, 0) = 
\begin{bmatrix}
    -r & -r & -aK \\
    0 & -\alpha & aK \\
    0 & \Omega\alpha & -aK - \delta
\end{bmatrix}.
\]
This matrix is of block upper triangular form. 
One eigenvalue is readily identified as \( \lambda_1 = -r < 0 \). 
The remaining eigenvalues are determined by the \( 2 \times 2 \) submatrix:
\[
A = 
\begin{bmatrix}
    -\alpha & aK \\
    \Omega\alpha & -aK - \delta
\end{bmatrix}.
\]
The characteristic equation for \( A \) is \( \det(A - \lambda I) = 0 \), which expands to:
\begin{equation}
    \lambda^2 + (\alpha + aK + \delta)\lambda + \alpha \left[ aK(1 - \Omega) + \delta \right] = 0. \quad
    \label{eq:A1}
\end{equation}
Applying the Routh-Hurwitz stability criterion to Eq.~(\ref{eq:A1}), the necessary and sufficient conditions for both eigenvalues to have negative real parts are:
(1)~\( \alpha + aK + \delta > 0 \) (always true for positive parameters),
(2)~\( \alpha \left[ aK(1 - \Omega) + \delta \right] > 0 \).
Given \( \alpha > 0 \) and noting that \( \Omega > 1 \), condition (2) simplifies to:
\[
\delta > aK(\Omega - 1).
\]
This inequality defines the critical adsorption rate:
\[
a_{\text{crit}} = \frac{\delta}{K(\Omega - 1)}.
\]
- If \( a < a_{\text{crit}} \), then \( \alpha \left[ aK(1 - \Omega) + \delta \right] > 0 \), all eigenvalues have negative real parts, and \( E_1 \) is locally, asymptotically stable.
- If \( a > a_{\text{crit}} \), then \( \alpha \left[ aK(1 - \Omega) + \delta \right] < 0 \), implying one positive real eigenvalue, and \( E_1 \) is unstable. 
This signifies that phage invasion is possible, and the system may evolve toward the coexistence equilibrium \( E_2 \).

The coexistence equilibrium satisfies \( \dot{B} = \dot{I} = \dot{P} = 0 \) with \( B^*, I^*, P^* > 0 \). 
Solving the equilibrium equations:
\[
\begin{aligned}
    0 &= rB^*\left(1 - \frac{B^* + I^*}{K}\right) - aB^*P^*, \\
    0 &= aB^*P^* - \alpha I^*, \\
    0 &= \Omega\alpha I^* - (aB^* + \delta)P^*,
\end{aligned}
\]
yields the biologically feasible solution:
\begin{equation}
    \begin{aligned}
        B^* &= \frac{\delta}{a(\Omega - 1)}\\
        I^* &= \frac{a}{\alpha}B^*P^*\\
        P^* &= \frac{r}{a}\left(1 - \frac{B^* + I^*}{K}\right)
        \label{eq:A2}
    \end{aligned}
\end{equation}

Note that \( B^* > 0 \) requires \( a > a_{\text{crit}} \), consistent with the instability condition for \( E_1 \). 
Substituting \( a_{\text{crit}} \) gives \( B^* = K (a_{\text{crit}} / a) \), confirming \( B^* < K \) only when \( a > a_{\text{crit}} \).

The stability of \( E_2 \) is determined by the eigenvalues of \( J(B^*, I^*, P^*) \). 
As the characteristic equation is cubic, the analytical evaluation of the real parts of the eigenvalues is complex. 
Numerical analysis of the maximum real part \( \Re(\lambda_{\text{max}}) \) gives the following stability regimes as parameters (e.g., the adsorption rate \( a \) or the growth rate \( r \)) vary:
(1) \( \Re(\lambda_{\text{max}}) < 0 \): \( E_2 \) is locally asymptotically stable (corresponding to the cyan region in the main text, Fig.~2~d).
(2) \( \Re(\lambda_{\text{max}}) = 0 \) and the imaginary part \( \Im(\lambda_{\text{max}}) \neq 0 \): A Hopf bifurcation occurs (indicated by dashed curves in Fig.~2~d)
(3) \( \Re(\lambda_{\text{max}}) > 0 \): \( E_2 \) is unstable, and the system undergoes a Hopf bifurcation, giving rise to a stable limit cycle (corresponding to the green region in Fig.~2~d). 
This results in sustained oscillations in the population densities of both bacteria and phages.

\section{Threshold-Induced Bistability and Basin of Attraction Geometry\label{app:basins}}

In the continuous deterministic framework, the extinction equilibrium $(0,0,0)$ is a saddle point with a one-dimensional unstable manifold along the $B$-axis and a two-dimensional stable manifold confined to the $I$--$P$ plane. 
Consequently, extinction is a measure-zero event: only trajectories initialized exactly on the stable manifold converge to $(0,0,0)$, while all others are repelled toward the coexistence attractor.

However, microbial populations are discrete and cannot recover from arbitrarily low densities. 
To account for this fact, we introduce a hard extinction threshold $\epsilon$, below which any population $X(t) \in \{B,I,P\}$ is set to zero and remains extinct. 
Mathematically, this implements an absorbing boundary at $X = \epsilon$. This seemingly simple modification fundamentally alters the phase space topology: the measure-zero stable manifold of the saddle point is effectively ``thickened'' into a finite-volume absorbing region $\mathcal{V}_{\text{ext}} = \{ (B,I,P) \mid \min(B,I,P) < \epsilon \}$. Trajectories that enter $\mathcal{V}_{\text{ext}}$ during transient dynamics are irreversibly captured by the extinction state, regardless of the linear stability of the coexistence equilibrium in the continuous limit.

Within parameter regimes where the coexistence state (or limit cycle) is linearly stable, the non-threshold system exhibits a single global attractor. 
The introduction of $\epsilon$ partitions the phase space into two basins of attraction: $\mathcal{B}_{\text{coex}}, \mathcal{B}_{\text{ext}}$. 
This constitutes a form of dynamical bistability induced not by multiple stable fixed points, but by the interplay between transient excursions and a finite-size absorbing boundary.

\begin{figure}[htbp]
\centering
\includegraphics[width=0.5\linewidth]{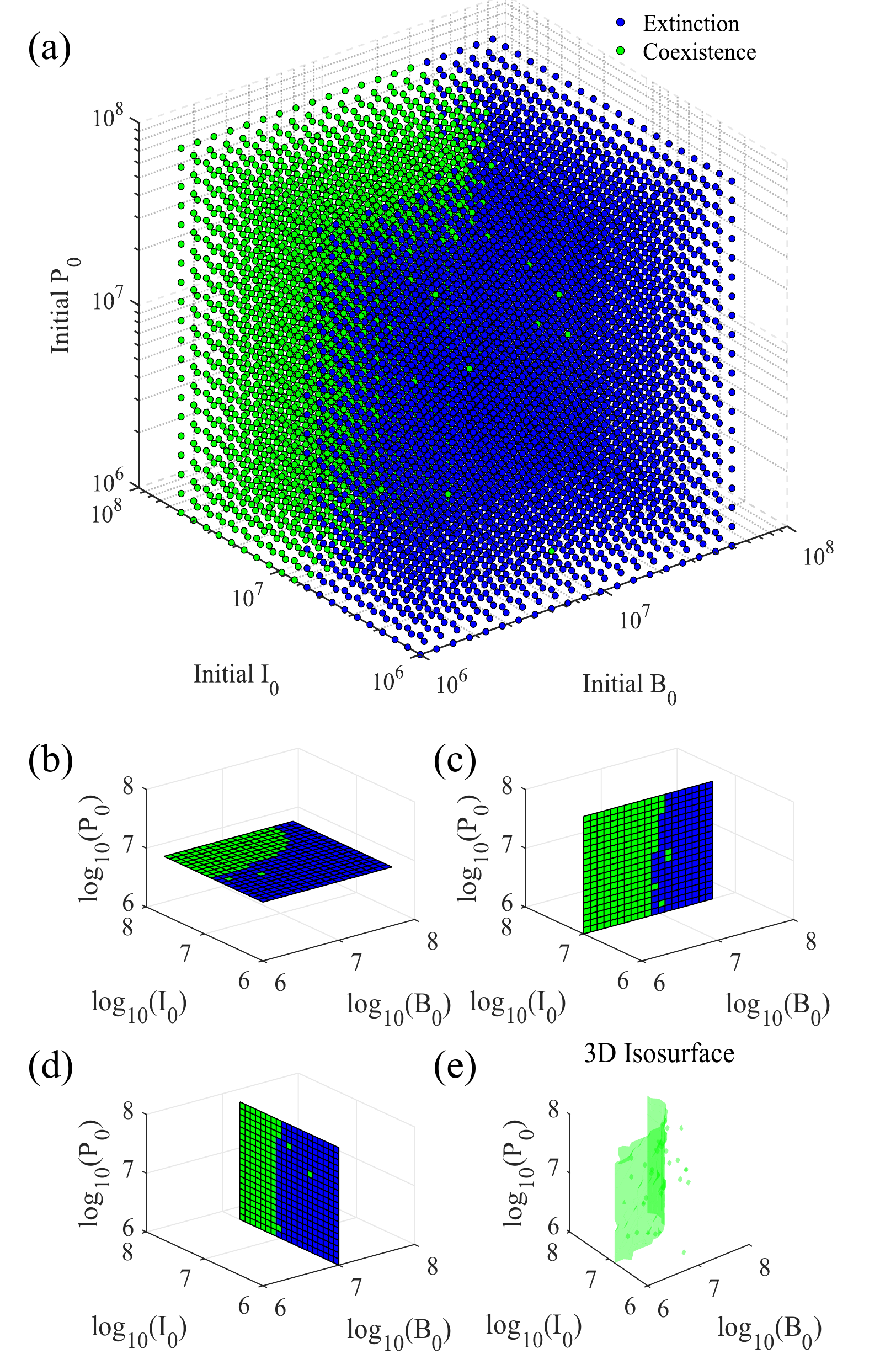}
\caption{\label{figS:basins}
Geometry of the basins of attraction under the extinction threshold.
(a) Three-dimensional scatter plot of initial conditions $(B_0, I_0, P_0)$, colored by outcome (green: coexistence/oscillations; blue: extinction). 
(b--d) Two-dimensional cross-sections along the $B_0$--$I_0$, $B_0$--$P_0$, and $I_0$--$P_0$ planes. 
(e) An iso-surface plot delineating the boundary between the basins of attraction for coexistence and extinction phase.
Parameters: $\hat{K}=1$, $\hat{a}=0.06$, $r=1$; extinction threshold $\epsilon=1$.}
\end{figure}

To quantitatively characterize this bistability, we map the basin of attraction for the coexistence state. 
We discretize the parameter space for the initial values of $B_0$, $I_0$ and $P_0$ on a logarithmic scale and perform long-term simulations for each combination. 
The resulting basin structure is visualized in Fig.~\ref{figS:basins}. 
Panel (a) presents the full three-dimensional distribution, revealing a non-negligible extinction basin (blue) embedded within the coexistence domain (green). 
Panels (b--d) show two-dimensional slices at fixed $P_0$, $I_0$, and $B_0$, respectively.

A feature of the basin geometry is the asymmetry in sensitivity across variables. 
As shown in Fig.~\ref{figS:basins}(c,d), varying $P_0$ over two orders of magnitude has negligible impact on the final outcome, whereas the $B_0$--$I_0$ plane exhibits a sharp transition boundary. 
We attribute this result to the rapid and explosive nature of the propagation of the phages. 
When the phages' burst size, $\Omega$, is sufficiently large, even vastly different initial values of $P_0$ will quickly converge to similar high level of phage densities during the initial outbreak stage.

\begin{figure}[htbp]
\centering
\includegraphics[width=0.8\linewidth]{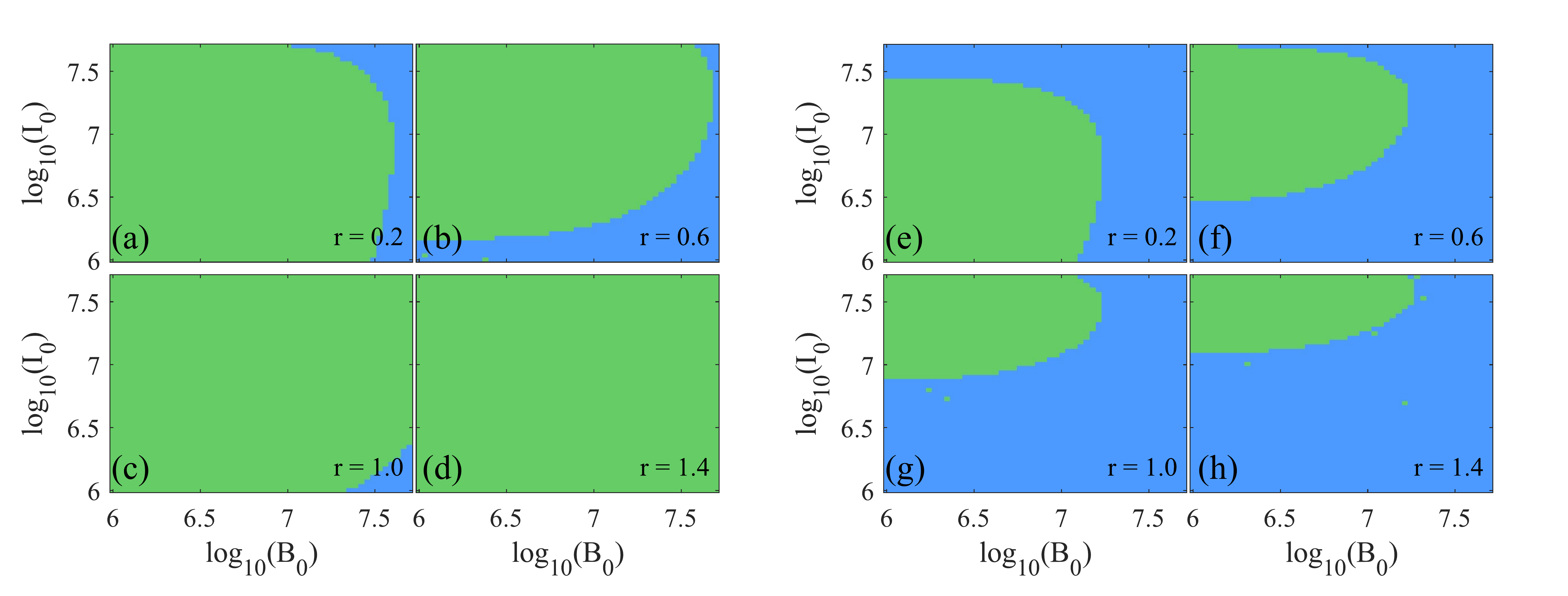}
\caption{\label{figS:survival_slices}
Cross-sections in the $B_0$--$I_0$ plane at fixed $P_0=0.1$. 
Green/blue regions denote initial conditions leading to coexistence/extinction. 
(a--d) Low adsorption rate ($\hat{a}=0.038$) with increasing $r$. 
(e--h) High adsorption rate ($\hat{a}=0.07$) with increasing $r$. 
The geometry of the survival basin is highly sensitive to both the bacterial growth rate and the adsorption rate.}
\end{figure}

The basin geometry is highly affected by host parameters. Figure~\ref{figS:survival_slices} illustrates how varying the bacterial growth rate $r$ and adsorption rate $a$ deforms $\mathcal{B}_{\text{coex}}$ in the $B_0$--$I_0$ plane. 
At low $a$, increasing $r$ expands the survival region. 
At high $a$, however, this trend reverses: higher $r$ amplifies transient oscillations shrinking the survival basin, while lower $r$ dampens oscillations and preserves a finite survival basin. 
This parameter-dependent change quantitatively shows the survival-probability reversal discussed in the main text, showing how host life-history traits shape the topological landscape of ecological persistence.

\section{Stochastic Simulation Validation of the Results}\label{app:Stochastic}
Stochastic simulations of the ODE model (Eq.~(1)) in this study were implemented based on the Gillespie algorithm. 
The state variable was defined as $\mathbf{X}(t) = [B(t), I(t), P(t)]^T$, where $B, I, P \in \{0, 1, 2, \dots\}$ represent the numbers of susceptible bacteria, infected bacteria and bacteriophages, respectively. 
The total bacterial population was given by $N = B + I$. 
Each term in the ODE model was decomposed into five reaction channels, and the propensity function corresponding to each reaction is summarized in Table~\ref{tab:stochastic_simulation}.

\begin{table}[htbp]
\centering
\caption{\label{tab:stochastic_simulation}%
Correspondence among reaction processes, corresponding ODE terms, propensity functions and state changes.
}
\scalebox{0.7}{
\begin{tabular}{lllll}
\toprule
No. & Reaction Process & Corresponding ODE Term & Propensity Function $a_\mu$ & State Change $(\Delta B, \Delta I, \Delta P)$ \\
\midrule
$R_1$ & Bacterial Birth & $r B$ & $a_1 = r \cdot B$ & $(+1, 0, 0)$ \\
$R_2$ & Bacterial Death by Competition & $-\frac{r}{K} N B$ & $a_2 = \frac{r}{K} \cdot (B+I) \cdot B$ & $(-1, 0, 0)$ \\
$R_3$ & Infection & $-a B P$ (and $+aBP$) & $a_3 = a \cdot B \cdot P$ & $(-1, +1, -1)$ \\
$R_4$ & Lysis & $-\alpha I$ (and $+\Omega \alpha I$) & $a_4 = \alpha \cdot I$ & $(0, -1, +\Omega)$ \\
$R_5$ & Phage Decay & $-\delta P$ & $a_5 = \delta \cdot P$ & $(0, 0, -1)$ \\
\bottomrule
\end{tabular}
}
\end{table}

We remark that, to facilitate the stochastic simulations, the carrying capacity $K$ was reduced to 1000, and the adsorption rate $a$ was correspondingly increased to maintain a reasonable effective adsorption rate $\hat{a}$. 
All other parameters remained identical to those listed in Table~I. 
Our simulation results indicate that this minor modification just leads to a slight deviation in the parameter regime, and all the qualitative properties of the results obtained under the parametrization in the main text are perfectly preserved.

\begin{figure}[htb]
    \centering
    \includegraphics[width=1.0\linewidth]{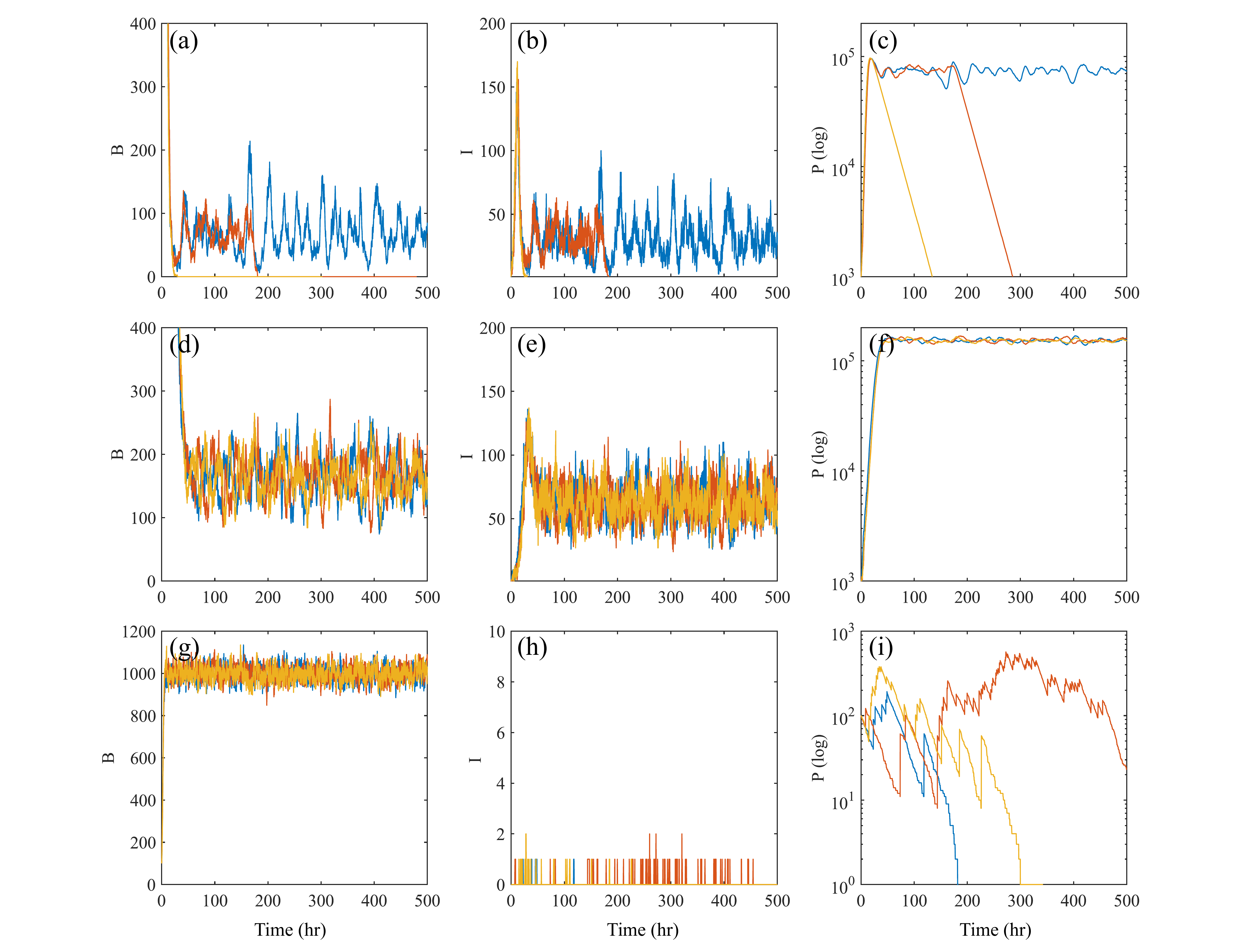}
    \caption{Stochastic dynamics of the bacteria-phage system across varying effective infection rates $\hat{a}$. 
    Rows correspond to $\hat{a}=0.012$, $0.005$, and $0.0009$ (top to bottom); columns show the trajectories of \textit{healthy} bacteria, \textit{infected} bacteria, and phages (left to right). 
    Each panel displays three independent stochastic realizations.}
    \label{figS:stochastic_dynamic}
\end{figure}

The results of stochastic simulations for the phage-bacteria dynamics are illustrated in Fig.~\ref{figS:stochastic_dynamic}. 
At high effective infection rates, bacteria exhibit pronounced oscillations that may lead to extinction. 
As $\hat{a}$ decreases, the oscillation amplitude diminishes and the average bacterial density increases, promoting a stable coexistence state. 
For extremely low infection rates, phages struggle to invade the host population, potentially resulting in their own extinction. 
These three dynamical regimes align qualitatively with those predicted by the deterministic ODE model.
As expected, unlike the fixed-frequency limit cycles in the ODE framework, the oscillation frequencies in the stochastic simulations exhibit intrinsic random fluctuations due to noisy effect.

\begin{figure}[ht!]
    \centering
    \includegraphics[width=0.8\linewidth]{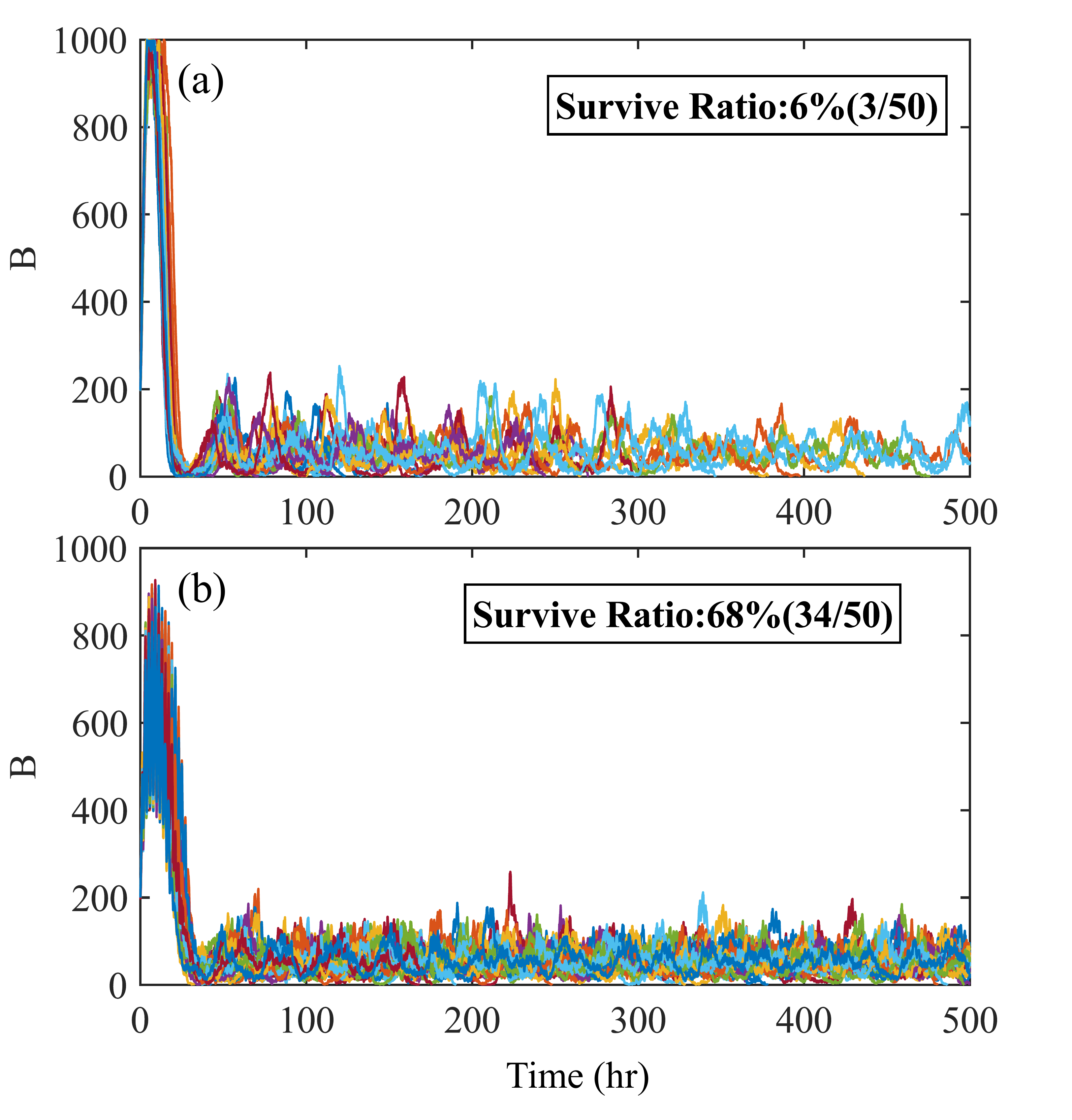}
    \caption{Comparison of bacterial survival under high infection pressure ($\hat{a}=0.015$) with and without periodic environmental fluctuations. (a) Time series of bacterial population from 50 independent stochastic simulations without environmental fluctuations. (b) Time series from 50 independent simulations with periodic sinusoidal environmental fluctuations (amplitude $A=800$, frequency $f=0.5$). The survival ratios are indicated in each panel. Environmental fluctuations substantially enhance bacterial survival probability under high infection pressure.}
    \label{figS:stochastic_survive}
\end{figure}

We further performed stochastic simulations to compare the survival ratios under high infection pressure with and without environmental fluctuations. The results, as shown in Fig.~\ref{figS:stochastic_survive}, demonstrate that despite the stochastic nature of the simulations, the survival ratio is significantly enhanced in the presence of environmental fluctuations. 
This provides solid validation of our main conclusion that environmental fluctuations can promote bacterial survival. 
A more detailed investigation of the role of periodic environmental fluctuations under stochastic dynamics is left for future work.

\section{Stochastic and Non-Sinusoidal Environmental Fluctuations}

In the main text, the environmental carrying capacity is modeled as a sinusoidal periodic function to capture the most common environmental oscillations, such as circadian rhythms, daily cycles in nutrient levels or temperature, etc. 
However, natural environments inevitably exhibit stochastic variability, and their temporal profiles are rarely strictly sinusoidal. 
To check the robustness of our core findings, we here examine how alternative forms of environmental forcing affect the long-term dynamics of our phage-bacteria system.

\begin{figure}[htbp]
    \centering
    \includegraphics[width=1.0\linewidth]{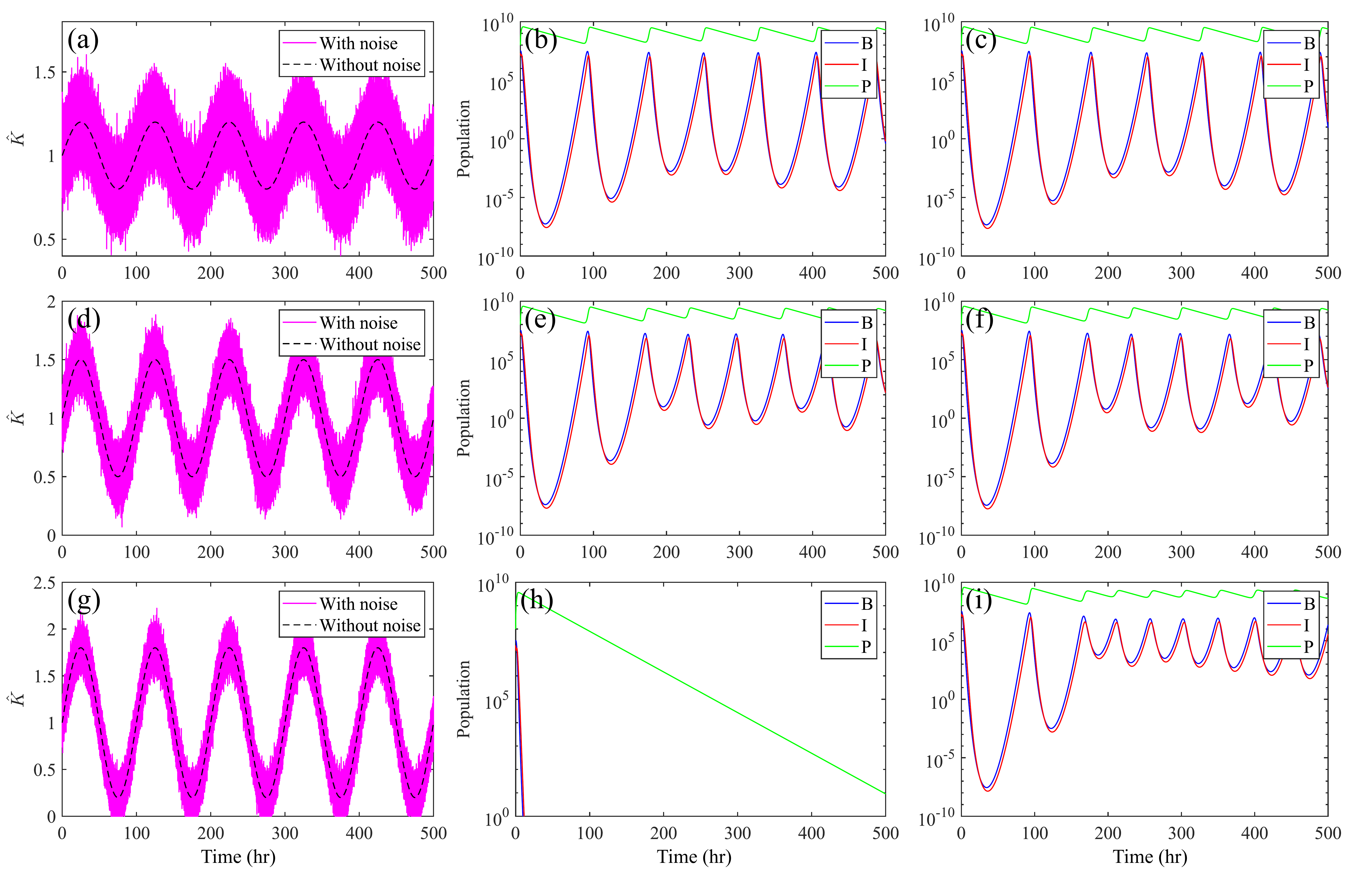}
    \caption{Population dynamics under sinusoidal carrying capacity modulated by Gaussian white noise.  
    Each row corresponds to a different modulation amplitude: $A=0.2$ (top), $A=0.5$ (middle), and $A=0.8$ (bottom), with a fixed noise intensity $\sigma_K=3.0$. 
    Left column displays two representative time series of the carrying capacity $K(t)$: the deterministic baseline and the stochastically perturbed realization. 
    Middle column shows the corresponding population dynamics driven by the noisy $K(t)$. 
    Right column presents the population dynamics under the deterministic (noise-free) $K(t)$ for direct comparison. 
    Despite the stochastic forcing, the qualitative dynamical regimes remain preserved across all amplitudes.}
    \label{figS:noise}
\end{figure}

First, we superimpose Gaussian white noise onto the baseline sinusoidal modulation. 
The carrying capacity is modeled as a stochastically perturbed periodic function:
\begin{equation}
    K(t) = \bar{K}(t) + \sigma_K \xi(t),
\end{equation}
where $\bar{K}(t) = K_0[1+A\sin(2\pi f t)]$ and $\xi(t)$ denotes Gaussian white noise with zero mean and delta-correlated unit intensity ($\langle \xi(t)\xi(t')\rangle = \delta(t-t')$). 
Due to the stochastic nature of the forcing, numerical simulations are performed using the Euler--Maruyama scheme. 
We compare the population dynamics driven by the noisy carrying capacity with those under the deterministic sinusoidal case, as shown in Fig.~\ref{figS:noise}. 
The results indicate that even relatively strong environmental white noise exerts only a marginal impact on the system's dynamical behavior. 
A pronounced deviation occurs only when the modulation amplitude $A$ is sufficiently large and stochastic excursions drive $K(t)$ toward zero, ultimately triggering population collapse. 
This effect is of physically intuitive and constitutes an expected consequence of severe resource depletion.

\begin{figure}[htbp]
    \centering
    \includegraphics[width=0.9\linewidth]{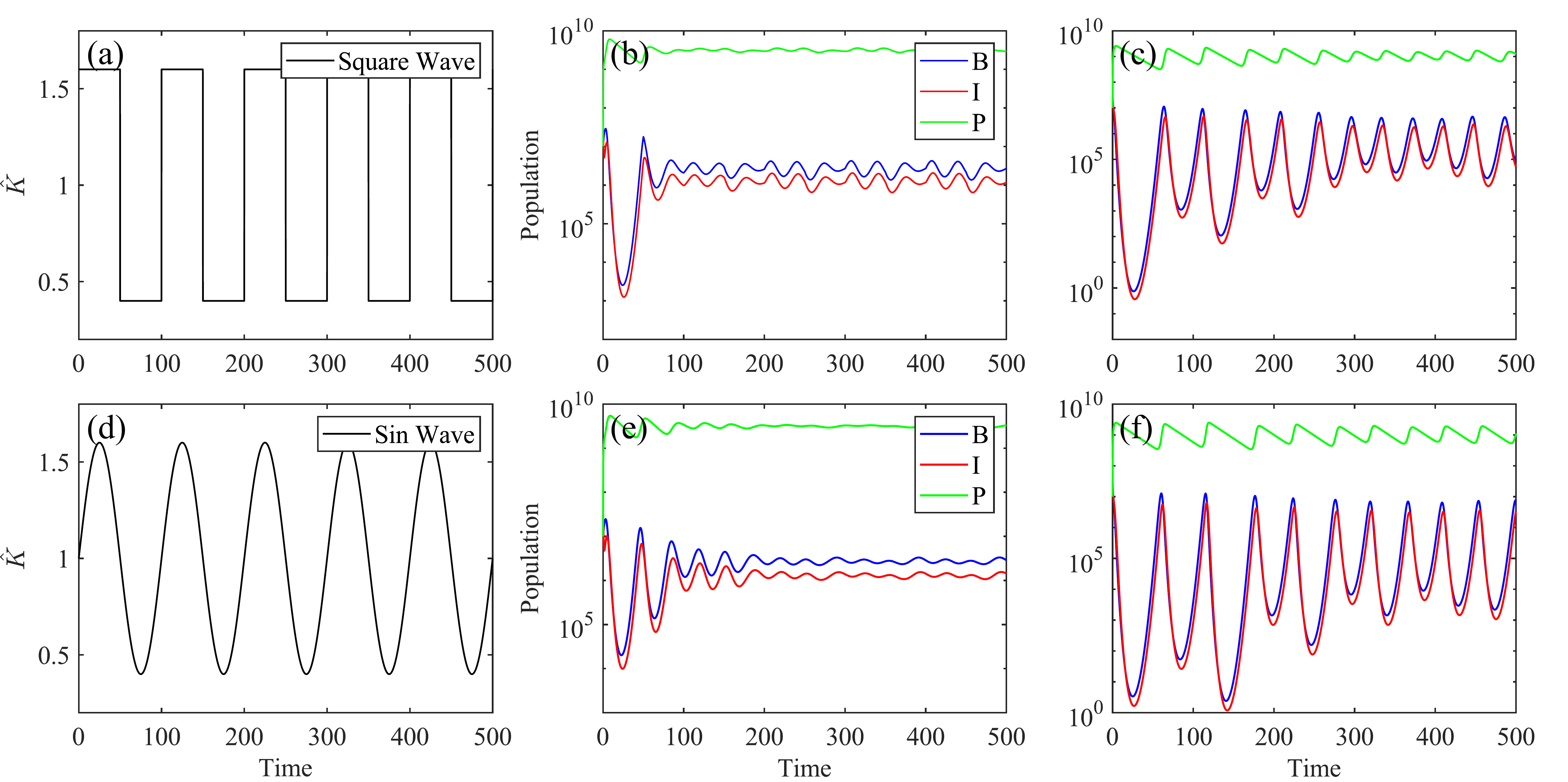}
    \caption{Comparison of population dynamics under square-wave versus sinusoidal environmental forcing. 
    The figure is arranged as a $2 \times 3$ grid. 
    Left column shows the time evolution of the carrying capacity $K(t)$ for a square-wave modulation (top row) and a sinusoidal modulation (bottom row). 
    Middle and right columns display the resulting population dynamics for two distinct baseline competition strengths: $\hat{a}=0.03$ (middle column), where the static system exhibits a stable coexistence equilibrium, and $\hat{a}=0.09$ (right column), where the static system operates near a limit cycle. 
    While the waveform change alters oscillation metrics quantitatively, the qualitative dynamical landscape is unchanged.}
    \label{figS:SquareWave}
\end{figure}

Next, we investigate non-sinusoidal periodic forcing, using a square-wave modulation as a representative example. 
Figure~\ref{figS:SquareWave} compares the population dynamics under square-wave and sinusoidal environmental forcing. 
Although the square-wave regime induces quantitative modifications in the frequency and amplitude of the population oscillations, the underlying qualitative dynamical behavior remains fully consistent with the sinusoidal case.

We find that neither stochastic perturbations of sinusoidal fluctuations nor alternative periodic waveforms qualitatively alter the stationary behavior of our studied system, thereby underscoring the robustness of our core results to a great extent. 
It should be noted, however, that our supplementary analysis is restricted to simple white noise and square-wave modulations. 
The potential impacts of more complex environmental variability--such as colored noise (e.g., Ornstein--Uhlenbeck processes) or multi-frequency superimposed fluctuations--still remain an open question and warrant further investigation in the future work.

\section{Supplementary Figures}

\begin{figure}[ht]
\centering
\includegraphics[width=0.6\linewidth]{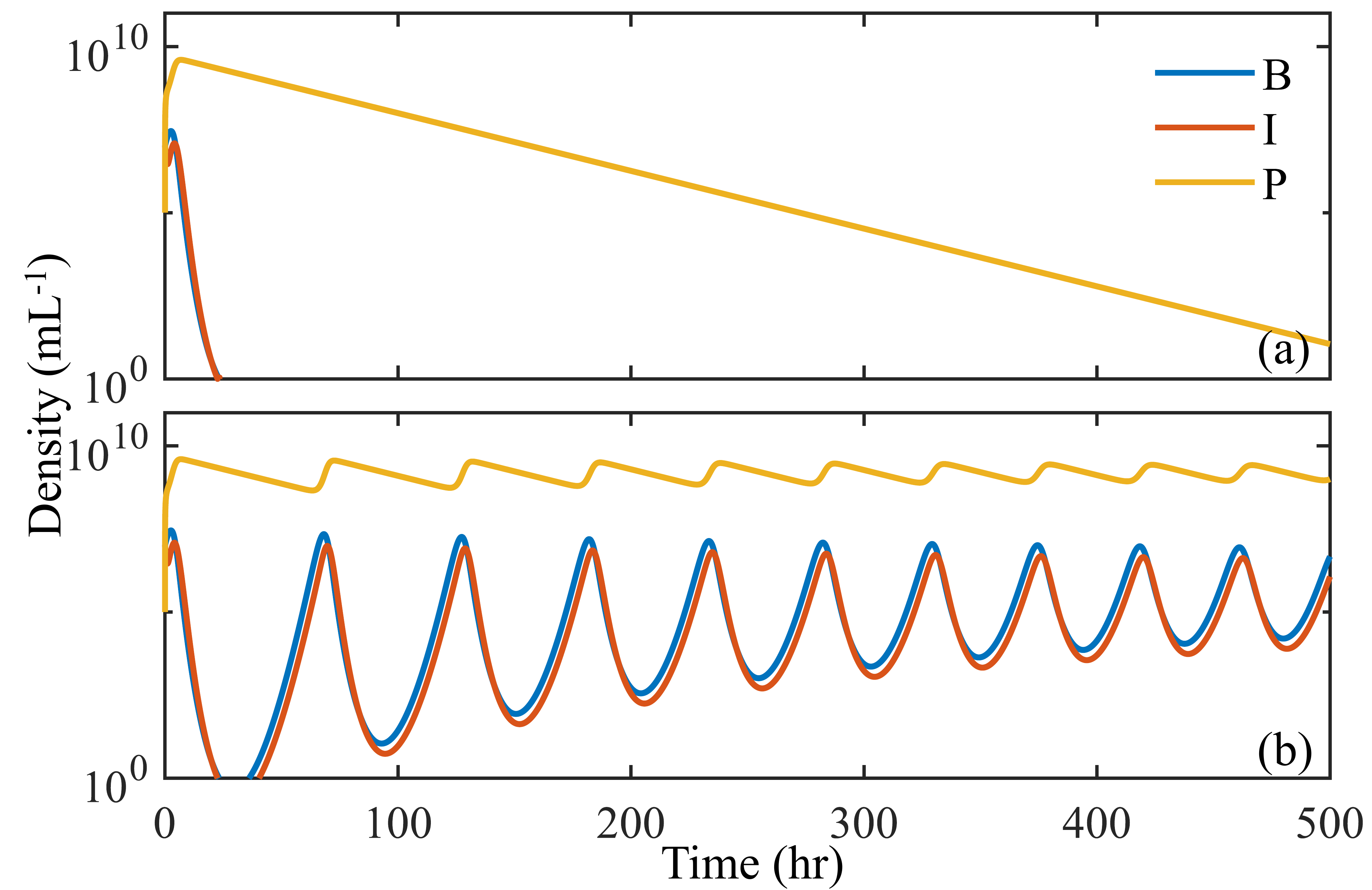}
\caption{\label{figS:timeseries1}Typical time series of population dynamics with and without an extinction threshold.}
\end{figure}
\noindent\textbf{Note on Fig.~\ref{figS:timeseries1}:} 
The panels compare temporal evolutions under identical parameters and initial conditions, differing only in the imposition of a demographic cutoff $\epsilon=1$. 
When the number of any species drops below $\epsilon$, it is set to zero [panel (a)]. 
This threshold can truncate transient oscillations and trigger a cascade leading to the eventual extinction of all populations. 
In the absence of the cutoff [panel (b)], the same trajectory sustains persistent (yet might be unphysical) oscillations. 
The comparison underscores how a realistic extinction threshold can qualitatively alter long-term fate during transient dynamics.

\begin{figure}[ht]
\centering
\includegraphics[width=0.6\linewidth]{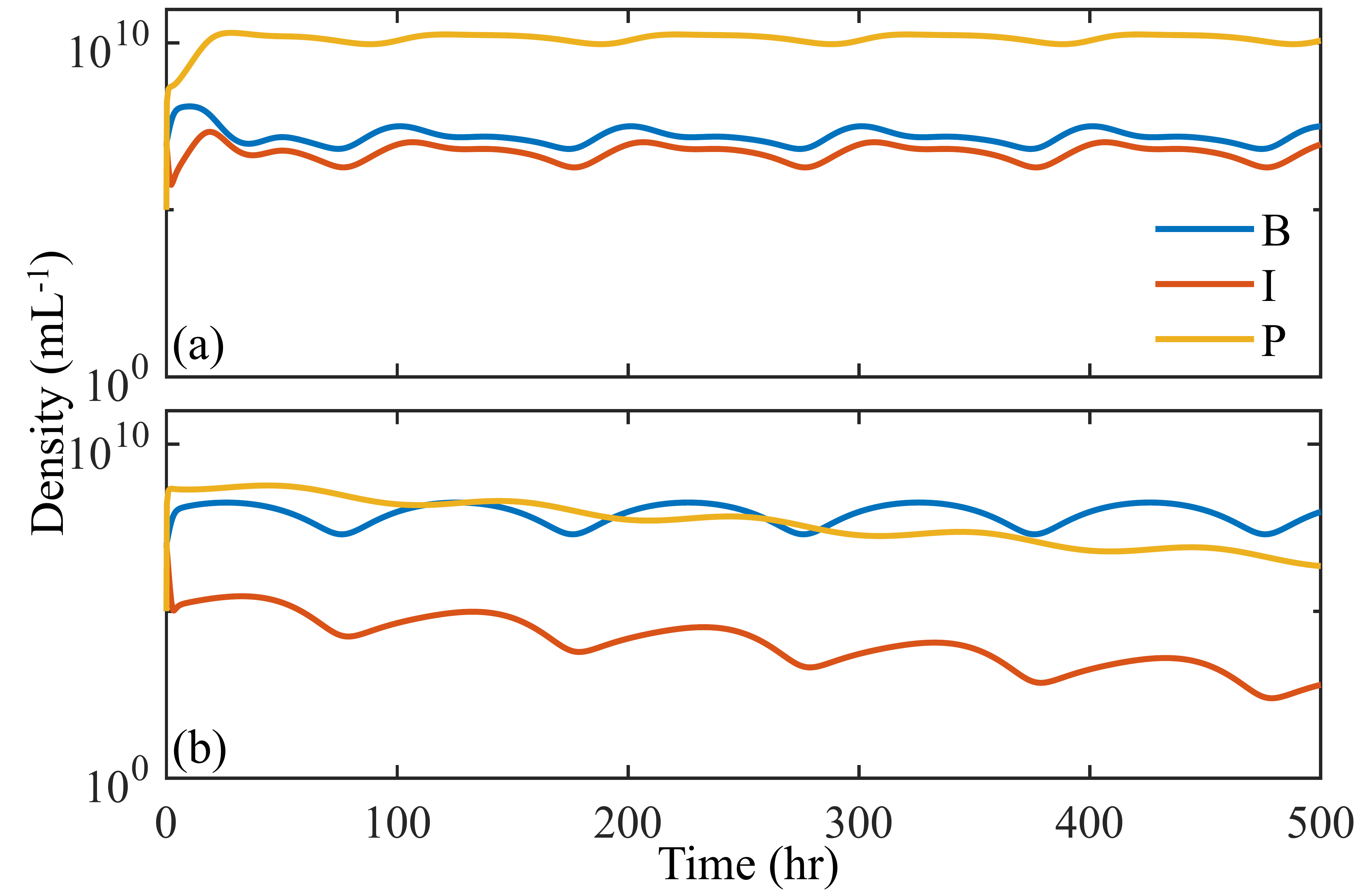}
\caption{\label{figS:timeseries2}Population dynamics under environmental fluctuations near deterministic fixed points.}
\end{figure}
\noindent\textbf{Note on Fig.~\ref{figS:timeseries2}:} 
Time series of the $B$, $I$, and $P$ populations are shown after introducing environmental stochasticity to previously stable deterministic states. 
Panel (a) corresponds to the coexistence fixed point (cyan region in Fig.~2d), while panel (b) corresponds to the bacteria-dominated fixed point (gray region in Fig.~2d). 
In both regimes, the system remains localized near its original state and exhibits no qualitative departure under moderate noise. 
Parameters: $A = 0.8$, $f = 0.01$; all other parameters are identical to those in Figs.~2b and 2c.

\begin{figure}[ht]
\centering
\includegraphics[width=1.0\linewidth]{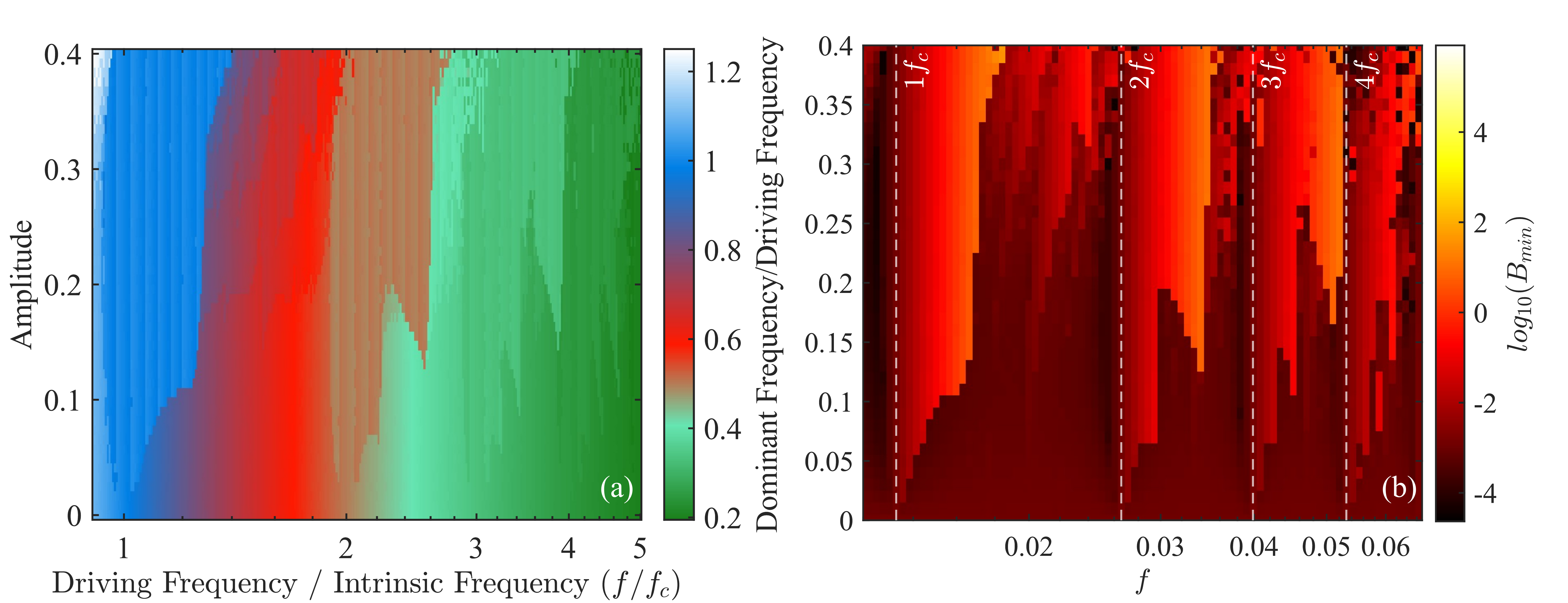}
\caption{\label{figS:Arnold}Enhanced visualizations of frequency locking and extinction risk.}
\end{figure}
\noindent\textbf{Note on Fig.~\ref{figS:Arnold}:} 
This figure provides optimized renditions of main-text Figs.~5 and~6 to clarify underlying structural features. Panel (a) reproduces the frequency-locking diagram with an adjusted color scale that sharpens the boundaries of the Arnold tongues. 
Panel (b) rescales the axes of the extinction-risk landscape to focus on the dynamically active parameter window. 
The side-by-side presentation makes the spatial alignment between the frequency-locked regions and the zones of suppressed extinction risk more explicitly, thereby visually confirming the mechanistic correspondence between phase synchronization and enhanced population resilience.

\end{document}